\documentclass[aps,prb,reprint,groupedaddress]{revtex4-2}

\usepackage{amsmath}
\usepackage{amsthm}
\usepackage{blindtext}
\usepackage{bm}
\usepackage{amsfonts}
\usepackage{graphicx}
\usepackage{amssymb}
\usepackage{feynmp}
\usepackage{breqn}
\usepackage{tikz}
\usepackage{tikz-feynman}
\usepackage[normalem]{ulem}

\newcommand{\avg}[1]{\langle #1 \rangle}

\newcommand{\kvec}{\mathbf{k}}
\newcommand{\qvec}{\mathbf{q}}

\newcommand{\Bvec}{\mathbf{B}}
\newcommand{\Tvec}{\mathbf{T}}
\newcommand{\Rvec}{\mathbf{R}}

\newcommand{\Hop}{\mathcal{H}}

\newcommand{\Sop}{\mathcal{S}}

\makeatletter
\let\cat@comma@active\@empty
\makeatother

\begin{document}


\title{Spin correlations in the bilayer Hubbard model with perpendicular electric field}


\author{Jiawei Yan}
\email{jiawei.yan@unifr.ch}
\affiliation{Department of Physics, University of Fribourg, 1700 Fribourg, Switzerland}
\author{Philipp Werner}
\email{philipp.werner@unifr.ch}
\affiliation{Department of Physics, University of Fribourg, 1700 Fribourg, Switzerland}


\date{\today}

\begin{abstract}
We present a nonequilibrium steady-state implementation of the two-particle self-consistent method. 
This approach respects the Mermin-Wagner theorem and incorporates non-local spatial fluctuations through self-consistent static vertices.
The real-frequency implementation allows to compute spectral properties without analytical continuation in both equilibrium and nonequilibrium.
As an interesting application, we investigate spin correlations in the bilayer square lattice Hubbard model under a perpendicular static electric field.
In equilibrium, the result yields spin correlations which are in good agreement with recent optical lattice experiments.
Under a large enough static electric field, the inter-layer spin correlations switch from antiferromagnetic to ferromagnetic.
We clarify how this phenomenon is linked to the nonequilibrium modifications of the spin excitation spectrum. 
\end{abstract}


\hyphenation{single}

\maketitle

\section{Introduction}\label{sec: introduction}

The Hubbard model is a minimal model for the description of electron correlation effects in solids. 
Despite its simplicity, the model exhibits rich equilibrium phase diagrams, which include magnetic phases, charge density waves and unconventional superconducting states \cite{Keimer2015, Galanakis2011}.
Recent advances in experimental techniques have extended the studies of correlated electron properties to the nonequilibrium domain. 
An interesting question in this context is the response of correlated electron systems to external electric fields \cite{Aoki2014,Murakami2023}. 
Unfortunately, due to the exponentially growing complexity with system size, the Hubbard model cannot be exactly solved in dimensions $D>1$, even in the single-band case.
The study of electric field effects introduces additional challenges, and necessitates the development of methods capable of addressing nonequilibirum conditions.

For equilibrium studies, powerful computational methods have been devised and used to reveal various Hubbard model properties \cite{LeBlanc2015, Qin2022, Schaefer2021, Orus2019}.
Dynamical mean-field theory (DMFT) \cite{Metzner1989,Georges1992} distinguishes itself by its nonperturbative nature, which makes it suitable for the study of strong correlation effects and Mott physics. 
Here, the lattice problem is mapped onto a single impurity coupled to a self-consistently computed non-interacting environment \cite{Georges1996,Kotliar2004}.
The effective impurity problem can be solved using numerically exact methods, such as quantum Monte Carlo \cite{Gull2011,Werner2006} or the numerical renormalization group  \cite{Wilson1975,Bulla2008}.
The local nature of the DMFT approximation however makes it unreliable in dimensions $D < 3$, where non-local spatial fluctuations become significant.
Although cluster or diagrammatic extensions of DMFT partially overcome this limitation, this comes at a considerable computational and storage cost \cite{Maier2005,Rohringer2018}. 
Nonequilibrium calculations have so far been limited to the dynamical cluster approximation (DCA) with small clusters \cite{Tsuji2014,Eckstein2016,Bittner2020}. Also, there is an unsolved conceptual issue how to properly incorporate electric fields into the DCA formalism. Consequently, there is an ongoing quest for novel methods to tackle nonequilibrium Hubbard systems.

Various semi-analytical methods capturing spatial correlations have been developed over the past decades \cite{Janis2008,Janis2017,*Janis2017b,Zalom2018,Vilk1997,Tremblay2012}.
These methods rely on renormalized perturbation theory \cite{Kadanoff1962, Bruus2004} and 
aim to describe correlation effects and critical behaviors by self-consistently calculating irreducible two-particle vertices.
This bottom-up approach allows them to mitigate spurious phase transitions that often occur in many-body perturbation theory (MBPT) and DMFT, when applied to low-dimensional systems. 
In contrast to the dynamical vertex approximation \cite{Held2008} or dual Fermion approach \cite{Rubtsov2008}, these methods assume spatially local and frequency-independent irreducible vertices.
These vertices are then determined self-consistently using either the reduced Parquet equation \cite{Janis2008,Janis2017,Zalom2018} or local sum rules \cite{Vilk1997,Tremblay2012}.
The former approach focuses on the qualitatively correct treatment across the entire parameter regime, while the latter aims to predict quantitatively accurate solutions in the weak-to-intermediate correlation regime (up to the regime where spin correlations start to become significant)
\cite{Vilk1997}.
In this work, we concentrate on the latter approach, specifically the two-particle self-consistent (TPSC) method, which is extended here to electric field driven nonequilibrium steady states. 

The TPSC method, originally introduced in Ref.~\cite{Vilk1994}, successfully reproduces the pseudo-gap associated with antiferromagnetic spin correlations in the square lattice Hubbard model.
The method is formulated in the thermodynamic limit, respects the Mermin-Wagner theorem, Pauli's exclusion principle, particle conservation laws and various sum rules \cite{Vilk1997,Tremblay2012}.
While originally designed for the single-band Hubbard model, the method has been extended in several directions, to take into account nearest-neighbor interactions \cite{Davoudi2007}, incorporate spin-orbital couplings \cite{Lessnich2023},  and multi-orbital interactions \cite{Miyahara2013,Zantout2021,GauvinNdiaye2023}.
Improved TPSC variants, named TPSC+ \cite{GauvinNdiaye2023a} or TPSC+GG \cite{Schaefer2021}, which (partially) feed back the spectral Green's function into the response function calculation, have also been developed. 
A recent addition is the combination of TPSC with DMFT, referred to as TPSC+DMFT, where the TPSC local self-energy is replaced by the DMFT self-energy \cite{Martin2023,Zantout2023,Simard2023}.
While these variants violate some sum rules, they enable a deeper exploration of the renormalized classical regime and (in the case of TPSC+DMFT) provide access to the Mott insulating regime. 
They also alleviate the overestimation of the spin correlations that occurs in the original theory, although this comes at the cost of smoothing out some features of the single-particle spectrum.
Moreover, TPSC and related methods are suitable for nonequilibrium extensions, since the underlying approximation on the irreducible vertex (completely local in time and space) significantly simplifies the implementation. 
Recently, the Anderson impurity model under external bias was studied using reduced Parquet equations \cite{Yan2022}, and a time-dependent TPSC formalism (with approximate solution of the Bethe-Salpeter equations) was implemented and applied to interaction quenches \cite{Simard2022,Simard2023}.

In this study, we demonstrate steady-state nonequilibrium extensions of the original TPSC method and of TPSC+GG.
In contrast to the time-dependent implementation \cite{Simard2022,Simard2023}, the Bethe-Salpeter equation can be exactly solved in frequency space, which ensures that the sum rules are strictly respected in our steady-state calculations.
The real-frequency implementation also avoids the large memory cost for storing two-time Green's functions, and it provides direct access to both equilibrium and nonequilibrium spectral properties without the need for numerical analytical continuation or (windowed) Fourier transformations.
As an application, we investigate a bilayer stack of the square lattice Hubbard model under a static perpendicular electric field.
Bilayer systems have been the focus of various recent studies, since the manipulation of bonding and antibonding states near the Fermi surface can lead to intriguing phenomena such as pairing correlations in the incipient bands  \cite{Karakuzu2021,Yue2022} and excitonic condensation \cite{Zeng2023}.
Notably, the response to a perpendicular electric field in the bilayer system is nontrivial, in contrast to the single-layer case, which predominantly exhibits a heating effect.
In the present study, we are primarily interested in the spin and charge correlation functions and their responses to the interlayer voltage bias. 

The rest of this article is structured as follows.
In Sec.~\ref{sec: theory}, we present the theoretical framework. Formally exact many-body equations are introduced in Sec.~\ref{subsec: many-body eqs}, followed by the TPSC approximation and formalism in Sec.~\ref{subsec: tpsc method}.
The implementation of our steady-state formalism is detailed in Sec.~\ref{subsec: ness implementatoin}.
In Sec.~\ref{sec: results}, the method is applied to the bilayer single-orbital Hubbard model.
We introduce the model and setup in Sec.~\ref{subsec: results - setup} and present the equilibrium and nonequilibrium results in Sec.~\ref{subsec: results - equilibrium} and \ref{subsec: results - nonequilibrium}, respectively.
In Sec.~\ref{sec: conclusions}, we give a brief conclusion, while detailed derivations and some TPSC+GG results are provided in the Appendices.

\section{Theory}\label{sec: theory}

We consider a single-band Hubbard model described by the Hamiltonian
\begin{dmath}\label{eq: Hamiltonian}
\Hop(t) = \sum_{ij}\sum_\sigma W_{ij}(t) c_{i\sigma}^\dag c_{j\sigma}
+ \sum_i U_i(t) \hat{n}_{i\uparrow} \hat{n}_{i\downarrow}~.
\end{dmath}
Here, $c_{i\sigma}^{(\dag)}$ represents the annihilation (creation) operator for site $i$ with spin $\sigma$, and $\hat{n}_{i\sigma} = c_{i\sigma}^\dag c_{i\sigma}$ denotes the corresponding density operator.
The on-site interaction is denoted by $U_i$ and the hopping amplitude from site $j$ to site $i$ by $W_{ij}$.
The chemical potential has been absorbed into the on-site components of the hopping matrix $W$.

\subsection{Many-body theory}\label{subsec: many-body eqs}

We start with formally exact equations from many-body perturbation theory (MBPT). 
A central quantity in this theory is the single-particle Green's function $G$, which is defined as
\begin{dmath}\label{eq: single-particle Green's function}
G_\sigma(1,2) = -i \avg{T_\gamma\{ c_\sigma(1), c_\sigma^\dag(2) \}}~,
\end{dmath}
where we adopt a compressed notation $1 \equiv (i,z)$, with $i$ and $z$ referring to the site and contour time, and $\sigma$ denoting spin. 
$T_\gamma$ is the time ordering operator on the contour $\gamma$, which, in the most general case, runs from time 0 to some maximum time $t_{\rm max}$ along the real axis, back to zero along the real axis, and then to $-i\beta$ (with $\beta=1/T$ the inverse temperature) along the imaginary-time axis \cite{stefanucci2013nonequilibrium}. (In the steady-state implementation, we employ a two-branch Schwinger-Keldysh contour \cite{Haug2008}.)
$T_\gamma$ orders the operators from right to left with increasing contour time, with an additional minus sign for the exchange of two Fermionic operators.
$\avg{\cdots} = \frac{1}{Z} \text{Tr} [ e^{-i\int_\gamma d\bar{z} \Hop(\bar{z})} \cdots ]$ denotes the expectation value, with $Z = \text{Tr} [ e^{-i\int_\gamma d\bar{z} \Hop(\bar{z})} ]$ the partition function of the initial equilibrium state.
Physically, $G_\sigma(1,2)$ represents the probability amplitudes for a single particle propagating in the interacting system.

By taking the derivative of Eq.~\eqref{eq: single-particle Green's function} with respect to time $z$ and considering the commutation relations of the operators, one obtains the equation of motion (EOM) of $G_\sigma(1,2)$, which in the case of the Hubbard model \eqref{eq: Hamiltonian} can be expressed as
\begin{dmath}\label{eq: eom of Green's function}
[\Sigma_\sigma * G_\sigma ](1,2) = -iU(1) \avg{T_\gamma\{ \hat{n}_{-\sigma}(1) c_\sigma(1), c_\sigma^\dag(2) \}}~.
\end{dmath}
Here, $\Sigma$ is the self-energy, which encodes how interactions affect the propagation of the electrons. $\Sigma_\sigma$ satisfies the Dyson equation
\begin{dmath}
G_\sigma(1,2) = G_\sigma^0(1,2) + [G_\sigma^0 * \Sigma_\sigma * G_\sigma](1,2)~,
\end{dmath}
 where $G^0_\sigma$ is the Green's function of the noninteracting model. 
The short-hand notation $[A*B](1,2) = \int_\gamma d\bar{3} A(1,\bar{3}) B(\bar{3},2)$ is used to denote convolutions over both the space and time domains.
Equation \eqref{eq: eom of Green's function} connects single-particle quantities (left hand side) to two-particle quantities (right hand side), and shows that in an interacting system, two-particle correlations affect the single-particle dynamics.

To quantify this, we introduce the generic four-point susceptibility
\begin{dmath}\label{eq: four-point susceptibility for hubbard model}
i\chi_{\sigma\sigma'}(1,2,3,4) = \avg{T_\gamma\{ c_\sigma(1) c_{\sigma'}(2) c_{\sigma'}^\dag(4) c_\sigma^\dag(3) \}} + G_\sigma(1,3)G_{\sigma'}(2,4)~,
\end{dmath}
which satisfies the Bethe-Salpeter equation (BSE)
\begin{dmath}\label{eq: BSE for hubbard model}
\chi_{\sigma\sigma'}(1,2,3,4) = G_\sigma(1,4)G_{\sigma'}(2,3) \delta_{\sigma\sigma'}-G_{\sigma}(1,\bar{1}) G_\sigma(\bar{3},3) \Lambda_{\sigma\bar{\sigma}}(\bar{1},\bar{2},\bar{3},\bar{4}) \chi_{\bar{\sigma}-\sigma}(\bar{4},2,\bar{2},4)~,
\end{dmath}
where $\Lambda_{\sigma\bar{\sigma}}(\bar{1},\bar{2},\bar{3},\bar{4})$ is the irreducible vertex.
We use the convention that variables with `overbars' are integrated over, and in the second term, we introduced the compact notation $\chi_{\sigma \sigma'}(1,2,3,4) \equiv \chi_{\sigma \sigma' \sigma \sigma'}(1,2,3,4)$.
Appendix \ref{sec: functional derivative technique} details the derivation of the above equations.
Combing Eqs.~\eqref{eq: eom of Green's function}, \eqref{eq: four-point susceptibility for hubbard model} and \eqref{eq: BSE for hubbard model}, we obtain the Schwinger-Dyson equation (SDE) for the self-energy:
\begin{dmath}\label{eq: SDE for hubbard model}
\Sigma_\sigma(1,2) = -iU(1) G_{-\sigma}(1,1^+) \delta(1-2) + U(1) G_\sigma(1,\bar{1}) \Lambda_{\sigma\bar{\sigma}}(\bar{1},\bar{2},2,\bar{4}) \chi_{\bar{\sigma}-\sigma}(\bar{4},1,\bar{2},1^+)~.
\end{dmath}
Here, $1^+ \equiv (i,z^+)$, where $z^+$ includes an infinitesimal time shift along the contour at $z$.

In the following, we focus on paramagnetic states, where $G_\sigma$ becomes spin-independent and $\chi_{\sigma\sigma'}$ can be transformed into contributions from spin and charge channels.
We define the spin and charge susceptibilities as $\chi^{\text{ch}} = \sum_{\sigma\sigma'} \chi_{\sigma\sigma'}$ and $\chi^{\text{sp}} = \sum_{\sigma\sigma'} \sigma\sigma' \chi_{\sigma\sigma'}$.
The physical two-point correlations are given by $\chi^{\text{sp}/\text{ch}}(1,2) = \chi^{\text{sp}/\text{ch}}(1,2,1^+,2^+)$, which can be expressed as
\begin{subequations}\label{eq: spin and charge response function}
\begin{align}
i\chi^{\text{sp}}(1,2)
&= \avg{\mathcal{T}_\gamma \{ \hat{S}^z(1), \hat{S}^z(2) \}}~, \\
i\chi^{\text{ch}}(1,2)
&= \avg{\mathcal{T}_\gamma\{ \hat{N}(1), \hat{N}(2) \}} - N(1) N(2)~.
\end{align}
\end{subequations}
Here, consistent with the naming convention, $\hat{N}(1) = \hat{n}_\uparrow(1) + \hat{n}_\downarrow(1)$ 
and $\hat{S}^z(1) = \hat{n}_\uparrow(1) - \hat{n}_\downarrow(1)$ are the total density and spin-$z$ operators (without factor $1/2$), respectively.
If one takes the limit $2 \rightarrow 1^+$ in Eq.~\eqref{eq: spin and charge response function}, one obtains the following local spin and charge sum rules \footnote{$2 \rightarrow 1^-$ leads to same results, since $\chi^{\text{sp/ch}}(1,1^+) = \chi^{\text{sp/ch}}(1,1^-)$.}:
\begin{subequations}\label{eq: local sum rules}
\begin{align}
i \chi^{\text{sp}}(1,1^+) &= N(1) - 2\avg{\hat{n}_\uparrow(1)\hat{n}_\downarrow(1)}~,\label{eq: local sum rules -- spin sector}\\
i\chi^{\text{ch}}(1,1^+) &=  2\avg{\hat{n}_\uparrow(1)\hat{n}_\downarrow(1)} -\left[N(1)\right]^2 + N(1)~. \label{eq: local sum rules -- charge sector}
\end{align}
\end{subequations}
where we used the property $(\hat{n}_\sigma)^2 = \hat{n}_\sigma$.

\subsection{TPSC method}\label{subsec: tpsc method}

The central approximation in TPSC is the following Hartree-type decomposition of Eq.~\eqref{eq: eom of Green's function}, 
\begin{dmath}\label{eq: Hatree decomposition}
[\Sigma^{(1)}_\sigma * G^{(1)}_\sigma](1,2) = -i \lambda(1) U(1) G^{(1)}_\sigma(1,2) G^{(1)}_{-\sigma}(1,1^+)~,
\end{dmath}
where $\lambda(1)$ is introduced as a renormalization factor to the bare Coulomb interaction.
The superscript `(1)' denotes the Green's function and self-energy defined in the first-level approximation.
The explicit expression of self-energy can be obtained by multiplying with $(G_\sigma)^{-1}$ from the right
\begin{dmath}\label{eq: thermodynamic self-energy}
\Sigma^{(1)}_\sigma(1,2) = -i\lambda(1) U(1) G^{(1)}_{-\sigma}(1,1^+) \delta(1-2)~.
\end{dmath}
If $\lambda(1)$ equals unity, this reduces to the conventional Hartree diagram.
The renormalized Coulomb interaction implies that the chemical potential should also be renormalized to give the correct filling.

The thermodynamically consistent irreducible vertex is obtained as $\Lambda_{\sigma\sigma'}(1,2,3,4) = -{\delta \Sigma^{(1)}_\sigma(1,3)}/{\delta G^{(1)}_{\sigma'}(4,2)}$, which yields
\begin{align}
&\Lambda_{\sigma\sigma'}(1,2,3,4)
= \nonumber\\
&\hspace{10mm} iU(1) \delta(1-3) \Bigg[ \lambda(1) \delta_{\sigma'-\sigma} \delta(1-4)\delta(1^+ -2) \nonumber\\
&\hspace{32mm} + \frac{\delta \lambda(1)}{\delta G^{(1)}_{\sigma'}(4,2) } G^{(1)}_{-\sigma}(1,1^+) \Bigg]~.
\end{align}
We furthermore assume that the second term in the brackets is proportional to $\delta(1-4)\delta(1^+ - 2)$,
to obtain the following local form of the spin and charge vertices
\begin{dmath}\label{eq: vertex expression}
\Lambda^{\text{sp/ch}}(1,2,3,4) = i\tilde{\Lambda}^{\text{sp/ch}}(1) \delta(1-3)\delta(1-4)\delta(1^+ -2)~,
\end{dmath}
where we defined $\tilde{\Lambda}^{\text{ch/sp}} = \tilde{\Lambda}_{\uparrow\downarrow} \pm \tilde{\Lambda}_{\uparrow\uparrow}$.
The factor $i$ in Eq.\eqref{eq: vertex expression} is chosen for convention in order to make $\tilde{\Lambda}(1)$ a real scalar.
One can easily see that
\begin{equation}\label{eq_gamma_spin}
\tilde{\Lambda}^{\text{sp}}(1) = \lambda(1) U(1)~,
\end{equation} 
since $\delta \lambda(1) / \delta G^{(1)}_\uparrow = \delta \lambda(1) / \delta G^{(1)}_\downarrow$ in the paramagnetic phase.

Given the completely local form of $\tilde{\Lambda}^{\text{sp/ch}}$, the BSEs can be simplified to
\begin{dmath}\label{eq: simplified BSE}
\chi^{\text{ch/sp}}(1,2) = \chi^0(1,2) \pm \frac{1}{2} \chi^0(1,\bar{1}) \tilde{\Lambda}^{\text{ch/sp}}(\bar{1}) \chi^{\text{ch/sp}}(\bar{1},2)~,
\end{dmath}
where $\chi^0$ in Eq.~\eqref{eq: simplified BSE} is the bare response function (also known as Lindhard function) given by
\begin{dmath}\label{eq: bare response function}
\chi^0(1,2) = -2i G^{(1)}(1,2) G^{(1)}(2,1)~.
\end{dmath}
Provided that the double occupancy is known, the sum rules in Eq.~\eqref{eq: local sum rules} allow to fix the local vertices $\tilde{\Lambda}^{\text{sp/ch}}$, since $\chi^{\text{sp/ch}}$ depend on $\tilde{\Lambda}^{\text{sp/ch}}$ via the BSEs. 
This is used for example in the recently developed TPSC+DMFT approach \cite{Martin2023}.

In the original TPSC, however, a further local field approximation (LFA) is introduced to self-consistently obtain the double occupancy without external input \cite{Vilk1994a}.
This is based on the fact that $\avg{\hat{n}_\uparrow(1)\hat{n}_\downarrow(1)}$ is spatially local and does not strongly depend on the nearby sites.
Specifically, in the hole doped case with density (per spin) $n$, the LFA makes the approximation
\begin{dmath}\label{eq: local field approximation}
\lambda(1) = \frac{\avg{\hat{n}^{(1)}_\uparrow(1) \hat{n}^{(1)}_\downarrow(1)}}{n^{(1)}(1) n^{(1)}(1)}~,
\end{dmath}
where $\avg{\hat{n}^{(1)}_\uparrow(1) \hat{n}^{(1)}_\downarrow(1)}$ denotes the double occupancy in the first-level approximation.
In the electron doped case, one should replace $\hat{n}_\sigma$ with $1 - \hat{n}_\sigma$ in Eq.~\eqref{eq: local field approximation} \cite{Vilk1997}.
With this approximation, the Hartree decomposition, Eq.~\eqref{eq: Hatree decomposition}, becomes identical to the EOM (Eq.~\eqref{eq: eom of Green's function}) in the limit $2\rightarrow 1^+$.
The self-consistent solution of Eqs.~\eqref{eq: local sum rules -- spin sector}, \eqref{eq: simplified BSE} and \eqref{eq: local field approximation} determines $\tilde{\Lambda}^{\text{sp}}$, and hence the double occupancy.
After fixing the double occupancy, one then searches $\tilde{\Lambda}^{\text{ch}}$ which satisfies Eq.~\eqref{eq: local sum rules -- charge sector}.
The above thermodynamically consistent two-particle self-consistency is also known as the first-level TPSC iteration.

Once the vertices $\tilde{\Lambda}^{\text{sp/ch}}$ and susceptibilities $\chi^{\text{sp/ch}}$ have been obtained in both the spin and charge channels, the spectral 
self-energy can be obtained by solving the Schwinger-Dyson equation, Eq.~\eqref{eq: SDE for hubbard model}.
Specifically, in the paramagnetic phase with local irreducible vertices, we have
\begin{dmath}\label{eq: spectral self-energy}
\Sigma^{(2)}(1,3) = -iU(1)G^{(1)}(1,1^+)\delta(1-3) + \Sigma^{C}[\alpha](1,3)~,
\end{dmath}
where the first term is the usual Hartree contribution and
\begin{dmath}\label{eq: correlation self-energy}
\Sigma^{C}[\alpha](1,3) = \frac{iU(1)}{8} G^{(1)}(1,3) \alpha(3) \left[ \tilde{\Lambda}^{\text{ch}}(3) \chi^{\text{ch}}(3,1) + 3\tilde{\Lambda}^{\text{sp}}(3) \chi^{\text{sp}}(3,1) \right]~,
\end{dmath}
is the correlation self-energy obtained by averaging the longitudinal and transversal channels \cite{Allen2004}.
Here, the superscript `(2)' denotes the second-level spectral quantities.
The coefficient $\alpha$ in Eq.~\eqref{eq: correlation self-energy}, which renormalizes the vertices, is determined by the local sum rules in the spectral calculation \cite{Vilk1997,Simard2022}, 
i.e., by the condition
\begin{dmath}
[\Sigma^{(2)}[\alpha]* G^{(2)}](1,1^+) = iU(1) \avg{\hat{n}^{(1)}_\uparrow(1) \hat{n}^{(1)}_\downarrow(1)}~.
\end{dmath}
Here, $G^{(2)}$ is the spectral Green's function corresponding to $\Sigma^{(2)}$.
The $\alpha$-renormalization has been found to be important for real-time simulations, possibly due to the approximation introduced in solving the BSEs \cite{Simard2022}. However, in the present steady-state implementation, we find that it has almost no effect on the results. We thus keep $\alpha = 1$ in all our calculations.

The self-consistent iteration contains the following steps: 
(i) Start with some initial guess for the double occupancy $\avg{\hat{n}^{(1)}_\uparrow\hat{n}^{(1)}_\downarrow}$ and calculate the corresponding $\tilde{\Lambda}^{\text{sp}}$ using Eqs.~\eqref{eq: local field approximation} and \eqref{eq_gamma_spin}.
(ii) Solve the spin channel BSE \eqref{eq: simplified BSE} to obtain $\chi^{\text{sp}}$.
(iii) Update the double occupancy, $\avg{\hat{n}^{(1)}_\uparrow\hat{n}^{(1)}_\downarrow}$, using Eq.~\eqref{eq: local sum rules -- spin sector}.
Repeat the procedures (i) - (iii) until convergence is reached in the spin channel.
(iv) Once the spin channel is converged, use the converged value of $\langle\hat{n}^{(1)}_\uparrow \hat{n}^{(1)}_\downarrow\rangle$ in the right-hand side of Eq.~\eqref{eq: local sum rules -- charge sector} and search for $\tilde{\Lambda}^{\text{ch}}$ that satisfies this sum rule.
After obtaining $\tilde{\Lambda}^{\text{sp/ch}}$ and $\chi^{\text{sp/ch}}$, one calculates $\Sigma^{(2)}$ using Eq.~\eqref{eq: spectral self-energy} and from this the corresponding $G^{(2)}$.

It is noteworthy that, for a generic system (such as the bi-layer lattice out-of-equilibrium, as discussed later), the iteration of $G^{(1)}(1,2)$ is essential to achieve first-level self-consistency. This is due to the fact that $\Sigma^{(1)}(1,2)$ in Eq.\eqref{eq: thermodynamic self-energy} is site-dependent and does not equate to a non-interacting system with a shifted chemical potential. Consequently, the Lindhard function (Eq.\eqref{eq: bare response function}) and Bethe-Salpeter equations need to be recalculated in each self-consistent loop.
In the case of $\Sigma^{(1)}(1,2)$ that is site-independent (for example single-band or bi-layer lattice at equilibrium), this self-consistency process can be simplified since the constant shift of the first-level self-energy can be absorbed into the global chemical potential.

Moreover, the original TPSC does not work deep inside the renormalized classical regime, where antiferromagnetic spin fluctuations dominate the physics \cite{Vilk1997}.
To gauge the validity of the results, in addition to comparing with numerically exact methods, one can perform internal accuracy checks by measuring the difference between $\text{Tr}\left[ \Sigma^{(2)}* G^{(1)} \right]$ and $\text{Tr}\left[ \Sigma^{(2)}* G^{(2)} \right]$, where the former one can be proven to be equal to $iU\avg{\hat{n}^{(1)}_\uparrow \hat{n}^{(1)}_\downarrow}$ \cite{Vilk1997} (see Appendix \ref{sec: tpsc consistency condition} for more details).
The relative error is defined by
\begin{dmath}\label{eq: internal error}
\text{rel. error} = \left| \frac{\text{Tr}[\Sigma^{(2)} * G^{(1)}] - \text{Tr}[\Sigma^{(2)} * G^{(2)}]}{\text{Tr}[\Sigma^{(2)} * G^{(1)}]} \right|~.
\end{dmath}

One of the recently developed improved TPSC variants, TPSC+GG, feeds the spectral Green's function $G^{(2)}$ back to $\chi^0$ in Eq.~\eqref{eq: bare response function}, thus reaching a self-consistency loop~\cite{Schaefer2021,Simard2023}.
Even though this approach in principle violates the local sum rules, numerical tests show an improvement of the two-particle correlation functions, compared to QMC results \cite{Schaefer2021}.

\subsection{Steady-state implementation}\label{subsec: ness implementatoin}

For the steady-state implementation of the formulas, we adopt a two-branch Schwinger-Keldysh contour, assuming that initial correlations of the system have been wiped out upon relaxation into a time-translation invariant steady-state.
We further assume the existence of translational invariance in the lattice model. 
The general procedure for the steady-state implementation involves obtaining the real-time equations by applying Langreth's rules \cite{stefanucci2013nonequilibrium, Haug2008} to the equations for the contour $\gamma$, followed by Fourier transformation of the relevant components to frequency and momentum space to facilitate the implementation \cite{Yan2022,Yan2023,Yan2016}.
This technique allows to simulate both thermal equilibrium states and nonequilibrium steady-states, and to obtain spectral functions without the need for numerical analytical continuation.

We use the vector $\Bvec$ to denote a site within the unit cell and $\Tvec$ to represent the translation vector between unit cells.
(In the case of the bilayer lattice considered in Sec.~\ref{sec: results}, the unit cell consists of $2$ sites.) 
We adopt the Fourier transform convention
$O(\omega; \kvec) = \sum_{\Rvec}\int_{-\infty}^\infty dt O_\Rvec(t) e^{i(\omega t - \kvec\cdot\Rvec)}$
and
$O_\Rvec(t) = \frac{1}{2\pi N_\kvec}\sum_{\kvec} \int_{-\infty}^\infty d\omega O(\omega; \kvec) e^{-i(\omega t - \kvec\cdot\Rvec)}$, where $N_\kvec$ is the number of $\kvec$-points (unit cells).
For simplicity, we omit the superscripts ${}^{(1)}$ and ${}^{(2)}$ in the following subsection.

\begin{widetext}
In frequency-momentum space, the  BSE for the spin (charge) channel, Eq.~\eqref{eq: simplified BSE}, reads
\begin{subequations}
\begin{dmath}
\chi^{\text{sp}(\text{ch}), r/a}_{\Bvec\Bvec'}(\omega;\qvec) = 
\chi^{0, r/a}_{\Bvec\Bvec'}(\omega;\qvec)
\mp \frac{1}{2} \sum_{\bar{\Bvec}}\chi^{0,r/a}_{\Bvec\bar{\Bvec}}(\omega;\qvec) \tilde{\Lambda}_{\bar{\Bvec}}^{\text{sp}(
\text{ch})} \chi_{\bar{\Bvec}\Bvec'}^{\text{sp}(
\text{ch}),r/a}(\omega;\qvec)~,
\end{dmath}
\begin{dmath}
\chi^{\text{sp}(
\text{ch}),\lessgtr}_{\Bvec\Bvec'}(\omega;\qvec) = \chi^{0,\lessgtr}_{\Bvec\Bvec'}(\omega;\qvec) \mp \frac{1}{2} \sum_{\bar{\Bvec}} \left[ \chi_{\Bvec\bar{\Bvec}}^{0,r}(\omega;\qvec) \tilde{\Lambda}_{\bar{\Bvec}}^{\text{sp}(\text{ch})} \chi_{\bar{\Bvec}\Bvec'}^{\text{sp}(\text{ch}),\lessgtr}(\omega;\qvec) + \chi_{\Bvec\bar{\Bvec}}^{0,\lessgtr}(\omega;\qvec) \tilde{\Lambda}_{\bar{\Bvec}}^{\text{sp}(\text{ch})} \chi_{\bar{\Bvec}\Bvec'}^{\text{sp}(\text{ch}),a}(\omega;\qvec) \right]~,
\end{dmath}
\end{subequations}
where the upper minus sign is for the spin channel and the lower plus sign for the charge channel, respectively, and the superscripts `$r$' (`$a$') refer to the retarded (advanced) components.
The Lindhard function \eqref{eq: bare response function} becomes
\begin{subequations}
\begin{dmath}
\chi_{\Bvec\Bvec'}^{0,r/a}(\omega;\qvec) = -\frac{i}{\pi} \frac{1}{N_\kvec} \sum_{\kvec} \int_{-\infty}^\infty dx
\left[ 
G_{\Bvec\Bvec'}^{r/a}(x;\kvec+\qvec) G^<_{\Bvec'\Bvec}(x-\omega;\kvec) + G_{\Bvec\Bvec'}^{<}(x;\kvec+\qvec) G^{a/r}_{\Bvec'\Bvec}(x-\omega;\kvec) \right]~,
\end{dmath}
\begin{dmath}
\chi_{\Bvec\Bvec'}^{0,\lessgtr}(\omega;\qvec) = -\frac{i}{\pi} \frac{1}{N_\kvec} \sum_{\kvec} \int_{-\infty}^{\infty} dx G_{\Bvec\Bvec'}^{\lessgtr}(x; \kvec+\qvec) G_{\Bvec'\Bvec}^{\gtrless}(x-\omega;\kvec)~.
\end{dmath}
\end{subequations}
The local spin and charge sum rules in Eq.~\eqref{eq: spin and charge response function} can be expressed as
\begin{subequations}
\begin{dmath}
\frac{1}{N_\kvec} \sum_{\qvec}\frac{1}{2\pi}\int_{-\infty}^\infty d\omega \chi_{\Bvec\Bvec}^{\text{sp},<}(\omega;\qvec) =
-i\left[ N_\Bvec - 2\avg{\hat{n}_{\Bvec\uparrow}\hat{n}_{\Bvec\downarrow}} \right]~,
\end{dmath}
\begin{dmath}
\frac{1}{N_\kvec} \sum_{\qvec}\frac{1}{2\pi}\int_{-\infty}^\infty d\omega \chi_{\Bvec\Bvec}^{\text{ch},<}(\omega;\qvec) = 
i\left[ N_\Bvec^2 - N_\Bvec - 2\avg{\hat{n}_{\Bvec\uparrow}\hat{n}_{\Bvec\downarrow}} \right]~.
\end{dmath}
\end{subequations}

For the spectral calculation, we use $\Sigma_{\Bvec\Bvec'}^{r/a}(\omega;\kvec) = U_\Bvec n_\Bvec \delta_{\Bvec\Bvec'} + \Sigma_{\Bvec\Bvec'}^{C,r/a}(\omega;\kvec)$ and $\Sigma_{\Bvec\Bvec'}^\lessgtr(\omega;\kvec) = \Sigma_{\Bvec\Bvec'}^{C,\lessgtr}(\omega;\kvec)$, where the self-energy contributions from electron correlations are given by
\begin{subequations}
\begin{dmath}
\Sigma_{\Bvec\Bvec'}^{C,r/a}(\omega;\kvec) =
+ \frac{i}{8 N_\kvec} \sum_{\qvec} \frac{1}{2\pi}\int_{-\infty}^\infty dx G_{\Bvec\Bvec'}^{r/a}(x;\kvec+\qvec) \alpha_{\Bvec'} \left[
\tilde{\Lambda}^{\text{ch}}_{\Bvec'} \chi^{\text{ch},<}_{\Bvec'\Bvec}(x-\omega;\qvec) U_\Bvec + 3\tilde{\Lambda}^{\text{sp}}_{\Bvec'}\chi^{\text{sp},<}_{\Bvec'\Bvec}(x-w;\qvec) U_\Bvec
\right]\\
+ \frac{i}{8 N_\kvec} \sum_{\qvec} \frac{1}{2\pi}\int_{-\infty}^\infty dx G_{\Bvec\Bvec'}^{<}(x;\kvec+\qvec) \alpha_{\Bvec'} \left[
\tilde{\Lambda}^{\text{ch}}_{\Bvec'} \chi^{\text{ch},a/r}_{\Bvec'\Bvec}(x-\omega;\qvec) U_\Bvec + 3\tilde{\Lambda}^{\text{sp}}_{\Bvec'}\chi^{\text{sp},a/r}_{\Bvec'\Bvec}(x-\omega;\qvec) U_\Bvec
\right]~,
\end{dmath}
\begin{dmath}
\Sigma_{\Bvec\Bvec'}^{C,\lessgtr}(\omega;\kvec) = \frac{i}{8 N_\kvec} \sum_{\qvec} \frac{1}{2\pi}\int_{-\infty}^\infty dx G_{\Bvec\Bvec'}^\lessgtr(x;\kvec+\qvec) \alpha_{\Bvec'} \left[ \tilde{\Lambda}_{\Bvec'}^{\text{ch}} \chi_{\Bvec'\Bvec}^{\text{ch},\gtrless}(x-\omega;\qvec)U_\Bvec + 3\tilde{\Lambda}_{\Bvec'}^{\text{sp}} \chi_{\Bvec'\Bvec}^{\text{sp},\gtrless}(x-\omega;\qvec)U_\Bvec \right]~.
\end{dmath}
\end{subequations}
\end{widetext}
Note that we have the symmetry relations $\chi^{r/a}_{\Bvec\Bvec'}(\omega; \kvec) = \chi^{a/r}_{\Bvec'\Bvec}(-\omega;-\kvec)$ and $\chi^{\lessgtr}_{\Bvec\Bvec'}(\omega;\kvec) = \chi^{\gtrless}_{\Bvec'\Bvec}(-\omega; -\kvec)$ both for $\chi^0$ and $\chi^{\text{sp/ch}}$.

The most time-consuming part in the evaluation of the above formulas is calculating the convolutions in frequency-momentum space, such as $C(\omega; \qvec) = \frac{1}{2\pi N_\kvec}\sum_\kvec \int_{-\infty}^\infty dx A(x; \kvec+\qvec) B(x-\omega; \kvec)$.
In our implementation, we employ fast Fourier transformation to the time-position domain, where the convolutions become point-wise multiplications. 
As a result, the computational complexity reduces from $N_\omega^2 N_\kvec^2$ to $N_\omega N_\kvec \log (N_\omega N_\kvec)$, where $N_\omega$ and $N_\kvec$ are the number of frequency and ${\bf k}$-points, respectively.
Additionally, since our implementation works with real frequencies, and the spectra may contains delta-peaks (for example, in the case of the thermodynamic Green's function $G^{(1)}$), a broadening of poles has to be introduced.
However, since our model will be coupled to external baths (see below), the smearing of the poles is taken care of by these baths.

As sketched in Fig.~\ref{fig: model}(a), the model is coupled to external leads, which are assumed to be in local equilibrium with chemical potential $\mu_i$ ($i=\text{top}$, bottom) and inverse temperature $\beta$. 
For $\mu_{\rm top}>\mu_{\rm bottom}$ the system is in a nonequilibrium steady state with a current flowing from the top to the bottom lead 
\cite{Datta2005}.
The effect of the leads is described by the lead self-energy, which depends on the detailed lead setup. 
Here, for simplicity, we adopt the wide-band limit (WBL), which assumes a flat density of states near the Fermi level with half-bandwidth $D$.
The explicit expression for the WBL lead self-energy (retarded component) for site $i$ and spin channel $\sigma$ is  
\begin{dmath}
\Sigma^{\text{ld},r}_{i\sigma}(\omega) = \frac{\Gamma_{i\sigma}}{\pi} \ln \left| \frac{D + (\omega - \mu_i)}{D - (\omega - \mu_i)} \right| - i\Gamma_{i\sigma} \theta(|\omega - \mu_i| < D)~,
\end{dmath}
where  $\Gamma_{i\sigma}$ is the coupling strength of the $i\sigma$ channel and $\theta$ is the Heaviside step function. 
The lesser part can be obtained from the fluctuation-dissipation theorem,
\begin{dmath}
\Sigma_{i\sigma}^{\text{ld},<}(\omega) = -2i f_\beta(\omega - \mu_i) \Im \Sigma_{i\sigma}^{\text{ld},r}(\omega)~,
\end{dmath}
with $f_\beta$ the Fermi function for inverse temperature $\beta$, since the leads are assumed to be in local equilibrium. 
Note that $\Sigma_{i\sigma}^{\rm ld}$ is site-diagonal in the WBL and that the total lead self-energy is the sum of $\Sigma_{i\sigma}^{\text{ld}}(\omega)$ over sites $i$.  
(Generically, the total lead self-energy could contain off-diagonal terms, which make it $\kvec$-dependent.) As follows from Fig.~\ref{fig: model}(a), the lead self-energy from the top (bottom) bath is added to the sites in the top (bottom) layer. 
The red and green ellipses in Fig.~\ref{fig: model}(a) mark two sites for measuring nearest-neighbor inter- and intra-layer correlations, as discussed in the following.

\section{Numerical results}\label{sec: results}

\subsection{Setup}\label{subsec: results - setup}

\begin{figure}
\includegraphics[width=1.0\linewidth]{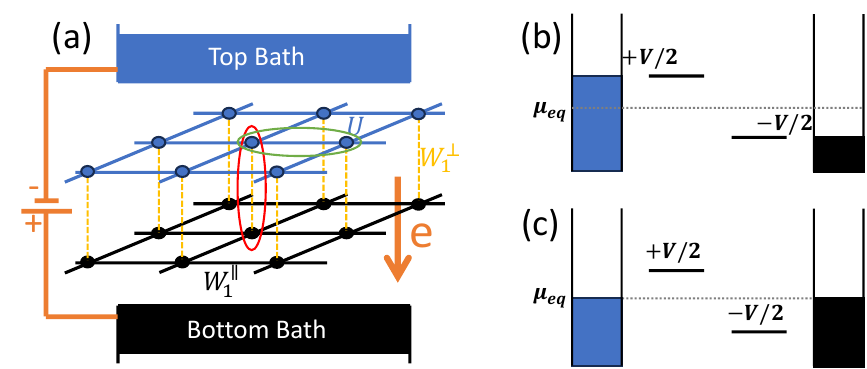}
\caption{(a) Schematic illustration of the bilayer square lattice Hubbard model with external voltage bias. Red and green ellipses mark nearest-neighbor inter- and intra-layer correlations, respectively. (b) Real space energy diagram for the nonequilibrium model, and (c) for the equilibrium model with local energy shifts.}\label{fig: model}
\end{figure}

We focus in the main text on a bilayer stack of the square lattice Hubbard model, with each layer coupled to a free-electron bath, as shown in Fig.~\ref{fig: model}(a).
The unit cell of the model comprises two sites and in the non-interacting case, the dispersion relation reads
\begin{dmath}\label{eq: bilayer dispersion relation}
\epsilon_{\pm}(\kvec) = 2 W^{\parallel}_1 \left[ \cos(k_x a) + \cos(k_y a) \right] \pm W_1^{\perp} + W_0~,
\end{dmath}
where $a = 1$ is the lattice constant, $W_{1}^{\parallel}$ and $W_{1}^{\perp}$ are the site-independent inter- and intra-layer hoppings and $W_{0}$ is the on-site energy, as depicted in Fig.~\ref{fig: model}. The plus and minus signs in Eq.~\eqref{eq: bilayer dispersion relation} refer to bonding (symmetric) and antibonding (antisymmetric) states resulting from the interlayer hopping.
One can also introduce a Fourier transform along the stacking direction, in which case the former (latter) state corresponds to $k_\perp = 0$ ($k_\perp = \pi$).

The baths are assumed to have an equilibrium distribution with inverse temperature $\beta$. An external applied voltage bias $V$ shifts their local energy levels and chemical potentials and results in a perpendicular electric field across the bilayer structure.
In the following study, we assume for simplicity that the voltage drop occurs only between the top and bottom layers, as depicted in Fig.~\ref{fig: model}(b). (A more advanced modeling would self-consistently determine the voltage profile by considering the charge density and Hartree potential.)
In our simple setup, the voltage bias has two main consequences: (i) a reshuffling of charge between the layers due to the different on-site energies of the two layers, and (ii) nonequilibrium distributions in the layers due to the different local chemical potentials in the baths.
To distinguish these two effects, we also consider an equilibrium model with shifted on-site energies of the layers, but with identical chemical potentials in the baths, see Fig.~\ref{fig: model}(c).

In the numerical calculations, we use a grid of $64 \times 64$ ${\bf k}$-points and $2^{16}$ frequency points on each side of the frequency axis, spaced by $d\omega = 10^{-3}$ to achieve a fine spectral resolution.
The computation time for each calculation is approximately 20 minutes when executed with 64 CPU cores.
\subsection{Thermal equilibrium}\label{subsec: results - equilibrium}

We first examine the equilibrium properties of the particle-hole symmetric system.
The parameters chosen in our study are $U=4$ and inverse temperature $\beta = 4$ (unless otherwise stated), and the coupling to each free-electron bath is $\Gamma=0.05$ (wide band approximation).
We set $W_1^{\parallel} = 1$ as the energy unit.

\begin{figure}
\includegraphics[width=1.0\linewidth]{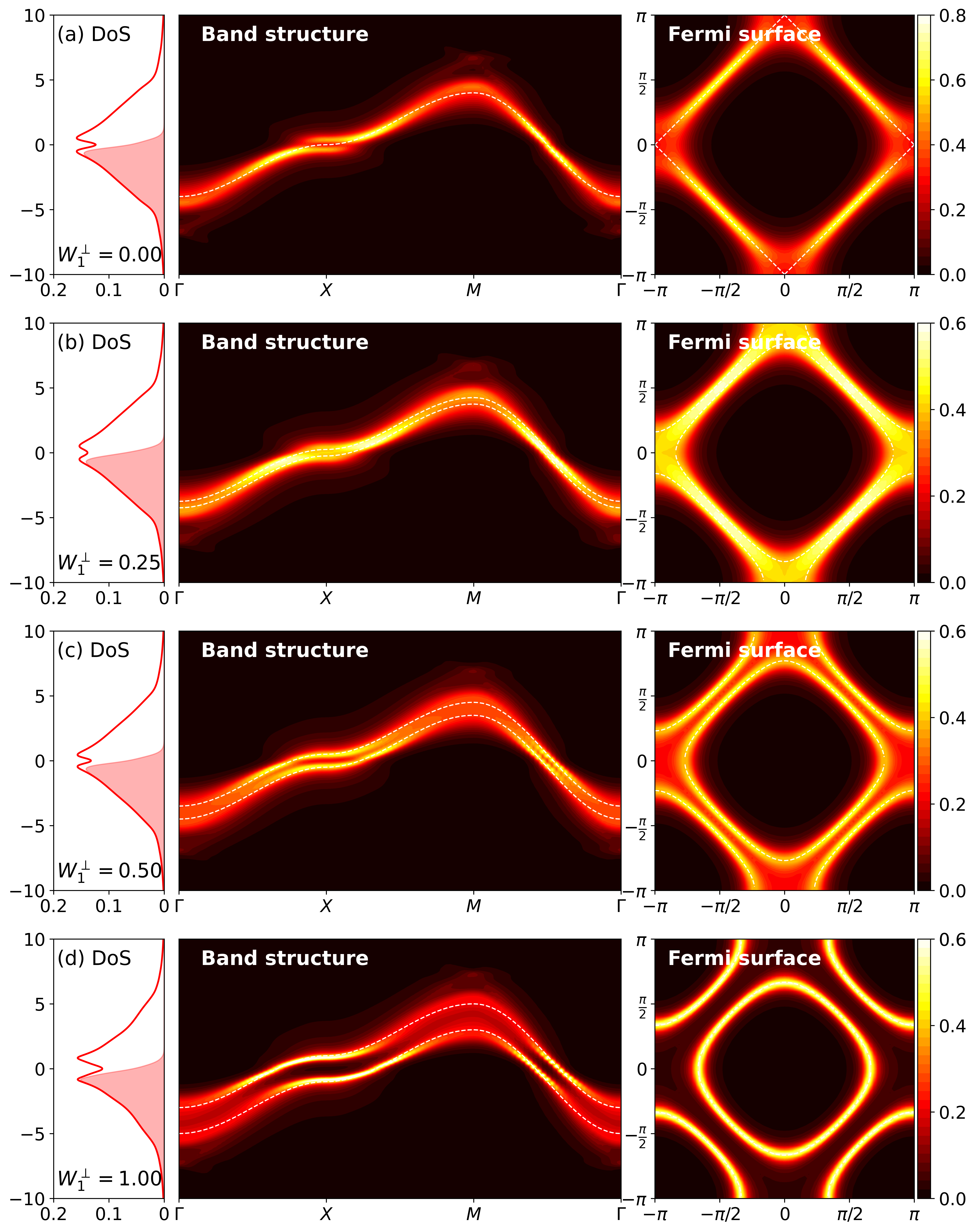}
\caption{Local spectral function $A(\omega)$ (shaded area for occupied states), momentum-resolved spectral function $A(\kvec; \omega)$ and ``Fermi surface" $A(\omega=0; \kvec)$ of the bilayer square lattice Hubbard model with (a) $W_1^{\perp} = 0$, (b) $W_1^{\perp} = 0.25$, (c) $W_1^{\perp} = 0.5$ and (d) $W_1^{\perp} = 1.0$, respectively.
Thin white dashed lines are the noninteracting bandstructures and Fermi surfaces. 
(Equilibrium system, $U=4$, $\beta=4$.)
\label{fig: tpsc_1Pqty_layer2_eq}}
\end{figure}

Figure~\ref{fig: tpsc_1Pqty_layer2_eq} presents the equilibrium ($\mu_{\text{top}} = \mu_{\text{bottom}}$) single particle quantities calculated by the original TPSC and for different inter-layer hoppings: $W_1^{\perp} = 0$ (a), $W_1^{\perp} = 0.25$ (b), $W_1^{\perp} = 0.5$ (c) and $W_1^{\perp} = 1.0$ (d).
From left to right, the three columns in Fig.~\ref{fig: tpsc_1Pqty_layer2_eq} display (i) the local spectral function $A(\omega)$ and occupation, (ii) the correlated band structure $A(\omega; \kvec)$ along a high symmetry path, $\Gamma=(0,0)$ -- $X=(\pi,0)$ -- $M=(\pi,\pi)$ -- $\Gamma=(0,0)$, and (iii) the ``Fermi surface" $A(\omega=0; \kvec)$.
The shaded area in the left panels marks the equilibrium distribution of occupied states, $A(\omega) f_\beta(\omega)$, where $f_\beta(\omega)$ is the Fermi function for inverse temperature $\beta$.

In the case $W_1^{\perp} = 0$, the two layers are decoupled and the model reduces to the single-layer square lattice model.
The corresponding local spectral function, shown in panel (a), features a pseudo-gap at the Fermi energy, indicative of strong antiferromagnetic correlations in the system. 
Indeed, the band structure and Fermi surface reveal a suppression of spectral weight at the $X$-point, $\kvec = (\pi,0)$, which results in a discontinuous Fermi surface.
The destruction of quasi-particles near the anti-node is consistent with the findings from other advanced many-body methods, such as cluster DMFT or dual Fermions  \cite{PhysRevResearch.2.033476, Brener2008}, but the TPSC approach is computationally much cheaper.
TPSC+GG results are presented in Appendix \ref{sec: single layer tpsc+gg}, where it is shown that this modified formalism does not produce an antiferromagnetic pseudo-gap in the spectral function for the same parameters. 

Panels (b), (c) and (d) in Fig.~\ref{fig: tpsc_1Pqty_layer2_eq} show how the electronic structure evolves as a function of the inter-layer hopping $W_1^{\perp}$.
One can see that the pseudo-gap at the Fermi level persists if $W_1^{\perp}$ is increased.
The underlying physics however changes, since the bilayer system exhibits a crossover from a pseudo-gap induced by antiferromagnetic correlations to a bonding/antibonding-type band splitting, as can be deduced from the band structures and Fermi surfaces \cite{Golor2014}.
As shown in the right panels, the Fermi surface represented by $A(\omega=0; \kvec)$ undergoes a nontrivial evolution: starting from a pseudo-gap pattern with disconnected segments, spectral weight redistribution leads to a splitting into two full Fermi surfaces with increasing $W_1^{\perp}$.

\begin{figure}
\includegraphics[width=1.0\linewidth]{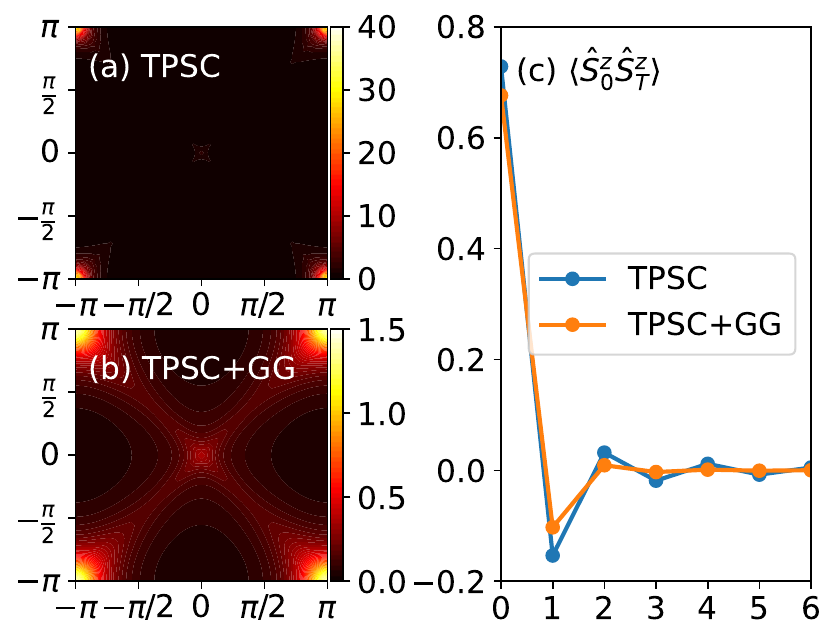}
\caption{(a) and (b) $-\frac{1}{\pi}\Im\chi^{\text{sp},>}(\omega=0; \qvec)$ in the first BZ obtained by TPSC and TPSC+GG, respectively. Note the different range of the color bars. (c) Corresponding real space spin-spin correlations $\avg{\hat{S}_0^z \hat{S}_T^z}$. The horizontal axis shows the distance (in units of lattice spacing) from the origin along the $x$ direction. (Equilibrium system, $U=4$, $\beta=4$.)
\label{fig: tpsc_2Pqty_eq}}
\end{figure}

In Fig.~\ref{fig: tpsc_2Pqty_eq}, we focus on the spin correlations.
Panels (a) and (b) depict the static spin susceptibility $-\frac{1}{\pi}\Im \chi^{\text{sp},>}(\qvec; \omega=0)$ obtained using TPSC and TPSC+GG, respectively.
One can see hot spots appearing at the corner $\qvec = (\pi,\pi)$ of the Brillouin zone (B), which confirms the existence of antiferromagnetic correlations in the system.
These correlations are strong in the TPSC solution, but much weaker in TPSC+GG (note the difference in the color bars), consistent with the presence/absence of a pseudo-gap in the fermionic spectral function noted above. 
In panel (c), we plot the corresponding real-space instantaneous ($t' \rightarrow t$) spin-spin correlations obtained by the Fourier transform 
\begin{dmath}\label{eq: real space spin-spin correlation formula}
\avg{\hat{S}^z_{\Bvec} \hat{S}^z_{\Bvec'+\Tvec}} = \frac{i}{2\pi N_\kvec} \sum_{\qvec} \int_{-\infty}^{\infty} d\omega \chi_{\Bvec\Bvec'}^{\text{sp},>}(\omega; \qvec) e^{-i\qvec\cdot\Tvec}~.
\end{dmath}
The $x$-axis corresponds to the distance (in units of lattice spacing) from the origin along the $x$ direction. By symmetry, the result is the same along the $y$ direction.
Due to the more smeared-out spin excitation spectrum, the TPSC+GG curve exhibits a faster decay compared to the TPSC result, indicating a shorter antiferromagnetic correlation length in the former method.

Previous studies have noted the overestimation of the spin correlation length by TPSC  \cite{Schaefer2021,Simard2023}, and it has been shown that TPSC+GG significantly improves the two-particle correlations. 
Based on this, one might conclude that the TPSC+GG results with shorter correlation length and nonexistent pseudo-gap should be more reliable for the present parameters.
On the other hand, it has been shown that self-consistent resummations via ``boldification" of diagrams can lead to inaccurate results \cite{Gukelberger2015} and in particular to unphysical fermionic spectral functions. 
A well-known example is the boldified second-order perturbation theory, which leads to a strong smearing of the spectral function and overdamped nonequilibrium dynamics \cite{Georges1992,Eckstein2010}.

\begin{figure}
\includegraphics[width=1.0\linewidth]{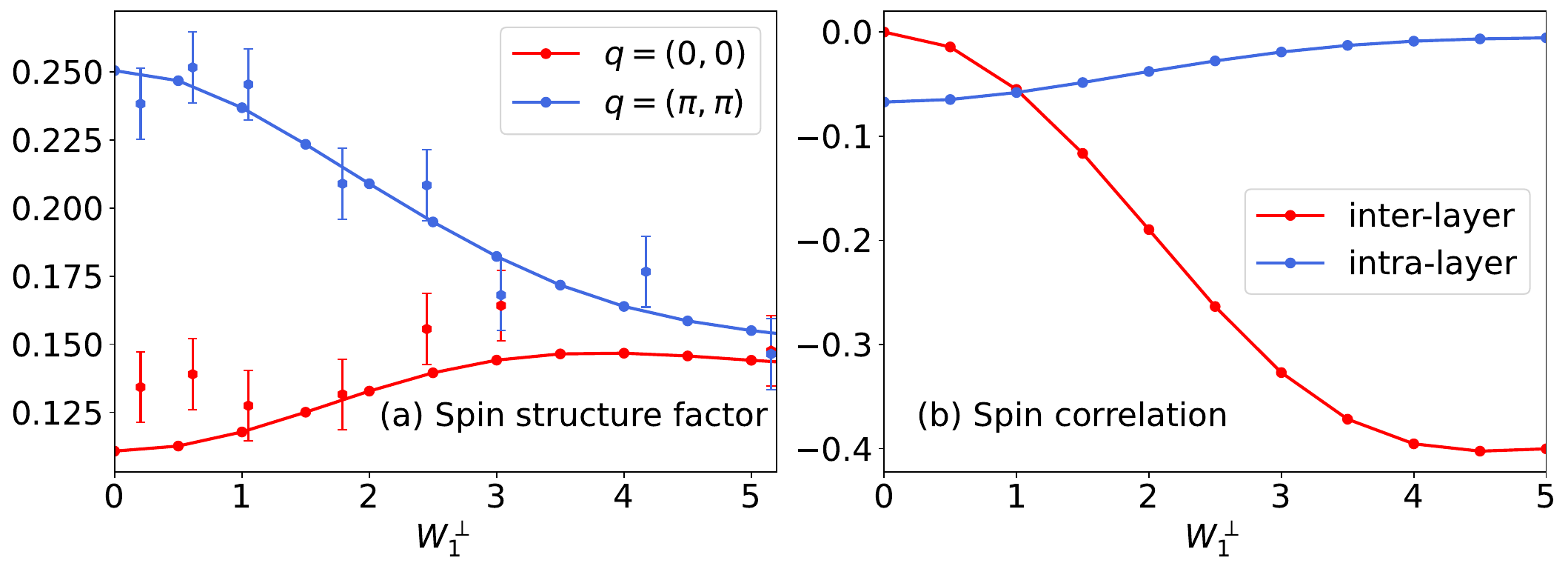}
\caption{(a) Dynamical spin structure factor (divided by $4$) for $\qvec=(0,0)$ and $\qvec=(\pi,\pi)$. The dots with error bars show the experimental data extracted from Ref.~\cite{Gall2021}. 
(b) Nearest-neighbor inter- and intra-layer spin-spin correlations. 
(Equilibrium system, $U=8$, $\beta=1$, density per spin $n=0.4$.)
\label{fig: benchmark_nature}}
\end{figure}

In the subsequent discussions, we primarily focus on the original TPSC method, which will be used to study the out-of-equilibrium behavior of the bilayer system.
Before that, we first benchmark the method using equilibrium data from a recent optical lattice experiment \cite{Gall2021}.
To ensure consistency with Ref.~\cite{Gall2021}, we adopt the same parameters $U=8$, $\beta = 1$ and $n = 0.4$ (average filling for each spin channel).
In Fig.~\ref{fig: benchmark_nature}(a), we plot the intra-layer (static) spin structure factor $i\chi^{\text{sp},>}_{00}(\omega=0;\qvec)$ as a function of inter-layer coupling $W_1^{\perp}$.
Results are shown for $\qvec = (0,0)$ (red color) and $\qvec = (\pi,\pi)$ (blue color).
The dots with error bars represent the experimental data extracted from Fig.~2(b) in Ref.~\cite{Gall2021}, where an optical lattice based quantum simulator was employed to study the bilayer Hubbard model.
Note that for the purpose of this comparison, the TPSC data are divided by $4$, since there is a factor $1/2$ difference in the definition of the spin operator in Ref.~\cite{Gall2021}.
The good agreement of the TPSC curves in Fig.~\ref{fig: benchmark_nature}(a) with the data from the optical lattice experiment provides support for the reliability of our method in the parameter regime considered in this study. 
In Fig.~\ref{fig: benchmark_nature}(b), we show the corresponding inter- and intra-layer spin correlations in red and blue, respectively.
As $W_1^\perp$ increases, the inter-layer spin correlations become negative, because the electrons on the different layers tend to form inter-layer singlets.
Simultaneously, the intra-layer nearest-neighbor spin correlations approach zero, since neighboring singlets become independent.

In all these equilibrium calculations, the internal error, as defined in Eq.~\eqref{eq: internal error}, remains below $10\%$. This indicates that the results in the chosen parameter regime do not have any obvious internal inconsistencies. 

\subsection{Nonequilibrium steady-state}\label{subsec: results - nonequilibrium}

We next apply a voltage bias $V$ across the bilayer system with $U=4$, to generate a static electric field perpendicular to the layers.
More specifically, the on-site energies of the top and bottom layers, as well as the respective baths, are shifted by $\pm V/2$ from their equilibrium values, as illustrated in Fig.~\ref{fig: model}(b).
As a result, electrons start to flow from the top to the bottom layers and the system reaches a steady-state (controlled by $V$ and the bath coupling $\Gamma$) after a sufficiently long time, which is the regime we analyze in the following.

\begin{figure}
\includegraphics[width=1.0\linewidth]{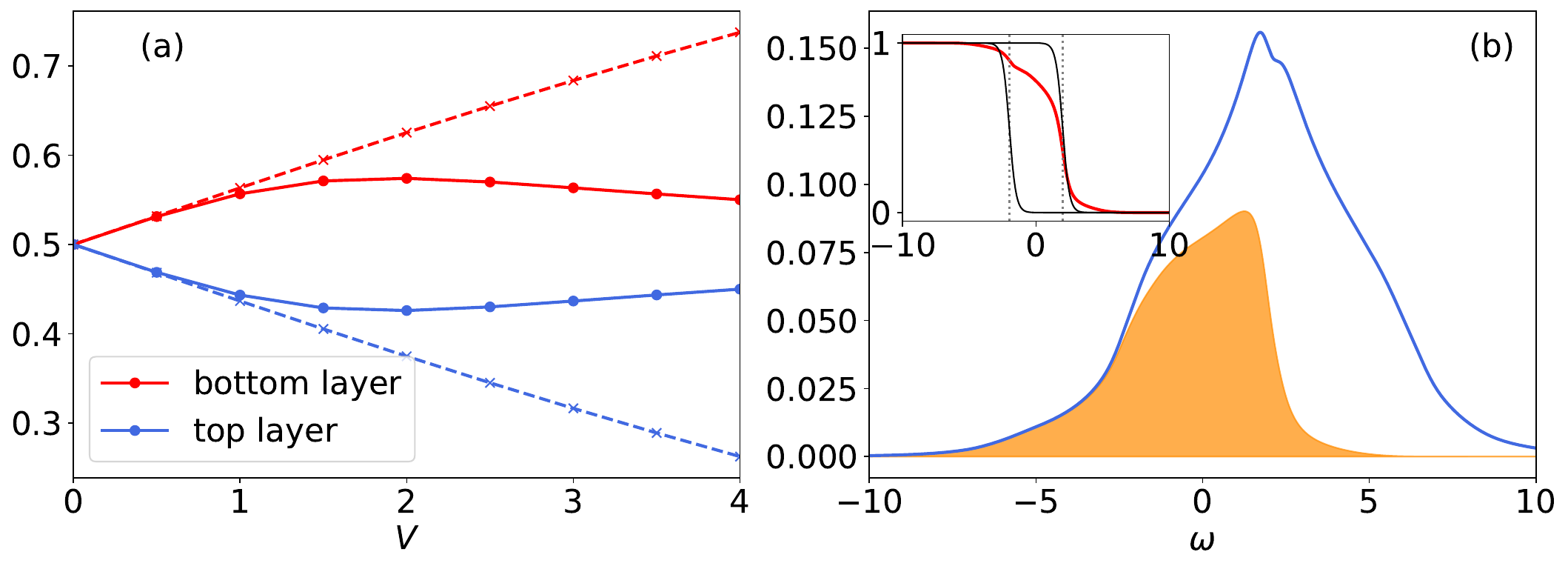}
\caption{(a) Electron density in the two layers versus voltage bias $V$. Solid and dashed lines show the results for the nonequilibrium setup, and the equilibrium setup with shifted local energies, respectively. (b) Spectral function of the top layer (blue line) $A_{\text{top}}(\omega)$ for $V=4$ and occupied states $A^<(\omega)$ (orange shading).
Inset: Corresponding nonequilibrium distribution function $f_{\rm neq}(\omega)$ (red). The black solid and grey dashed lines show the bath distribution functions and chemical potentials, respectively. 
(Parameters: $W_1^{\perp}=1$, $U=4$, $\beta=4$.)
\label{fig: bilayer_neq_1p}}
\end{figure}

\begin{figure*}[ht]
\includegraphics[width=1.0\linewidth]{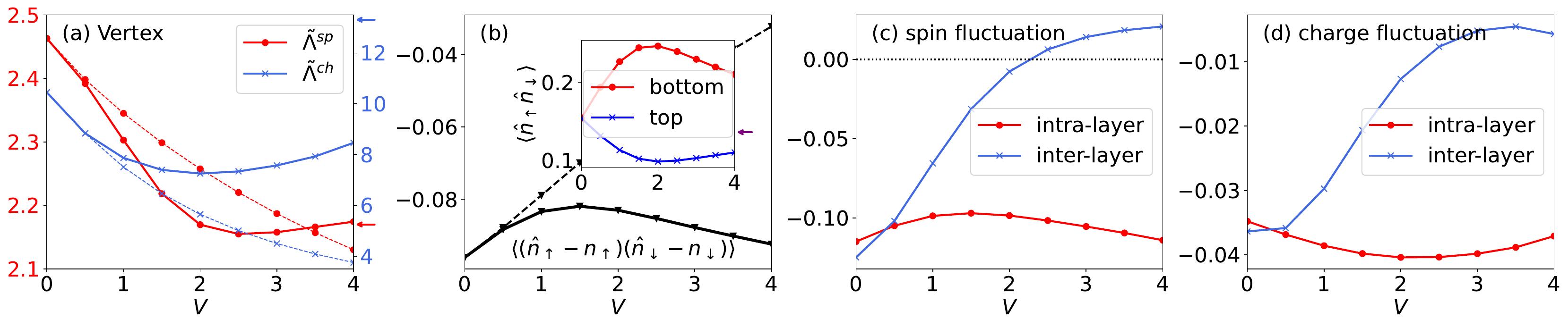}
\caption{Bias dependent two-particle quantities for $W_1^{\perp}=1$, $U=4$ and $\beta=4$. (a) Spin (red) and charge (blue) vertices. (b) Double occupancy fluctuation $\Delta D=\avg{({\hat{n}}_\uparrow - n_\uparrow ) ( {\hat{n}}_\downarrow - n_\downarrow )}$, which is the same for both layers. Inset: layer dependent $\avg{\hat{n}_\uparrow \hat{n}_\downarrow}$. In panels (a) and (b), the dashed lines show the results for the equilibrium model with shifted local energies and the arrows on the right indicate the asymptotic values. (c,d) Nearest-neighbor intra- and inter-layer spin and charge fluctuations, respectively.
\label{fig: bilayer_2p}} 
\end{figure*}

\subsubsection{Single-particle properties}

We start with the single-particle properties plotted in Fig.~\ref{fig: bilayer_neq_1p} for $W_1^{\perp} = 1$.
Panel (a) shows how the electron density (red for the bottom layer and blue for the top layer) varies as a function of the voltage bias.
Solid lines plot the densities calculated for the nonequilibrium model, while the dashed lines are the results for the equilibrium model with shifted on-site energies (see Fig.~\ref{fig: model}). 
With increasing $V$, there is a reshuffling of charge from the top to the bottom layer. (We neglect the effect of this reshuffling on the electric field or voltage drop within the bilayer.) 
When $V < 1$, the solid and dashed lines are close to each other, which demonstrates that in this regime, the charge reshuffling is dominated by the energy level shift.
For larger $V$, the decrease (increase) of the electron density in the top (bottom) layer of the nonequilibrium system becomes slower and after $V=2$, the trend is reversed.
Physically, this happens because the overlap in the densities of states of the top and bottom layer is reduced with increasing $V$, so that the transfer of electrons between the two layers slows down. Note that in the limit of no hopping between the layers, both layers will be half-filled in our nonequilibrium setup. 

In panel (b), we plot, in blue, the local spectral function $A(\omega) = - \Im G^r(\omega) / \pi$ of the top layer. (The spectral function for the bottom layer is mirrored at $\omega=0$.) 
The occupied states $A^<(\omega) = \Im G^<(\omega)/ (2\pi)$ of the upper layer are shown by the orange shading. The ratio between these two spectra defines the nonequilibrium distribution function $f_{\text{neq}}(\omega)= A^<(\omega) / A(\omega)$, which is plotted as a red curve in the inset. 
This nonequilibrium distribution function deviates significantly from the Fermi function of the upper bath (right black line), and gives rise to the interesting nonequilibrium phenomena discussed below.

\subsubsection{Two-particle properties}

We now turn to the investigation of two-particle quantities, fixing $W_1^{\perp} = 1$ and $\beta = 4$ unless otherwise stated. 
Panel (a) of Fig.~\ref{fig: bilayer_2p} plots the spin (red) and charge (blue) irreducible vertices as a function of the voltage bias.
Again, the solid and dashed lines show the results for the nonequilibrium setup and the equilibrium system with shifted local energies, respectively.
Due to the electron-hole symmetry in the structure, the local vertices are identical on both layers.
In contrast to the single layer case (see Appendix \ref{sec: single layer heating}), where an increase of $V$ mainly leads to a heating effect, the vertices in the bilayer system exhibit a non-monotonic behavior.
Specifically, for small bias, both $\tilde{\Lambda}^{\text{sp}}$ and $\tilde{\Lambda}^{\text{ch}}$ decrease with increasing $V$.
This effect mainly comes from the change in the filling of the layers, as the equilibrium model with shifted local energies predicts the qualitatively same behavior at small $V$.
For $V\gtrsim 2$, both the spin and charge vertices start to grow. 
As discussed above, this is because for large $V$, the layers are getting decoupled (reduced overlap in the local densities of states), so that the vertices are expected to approach the equilibrium values for decoupled layers with $\beta=4$.
These values, $\tilde{\Lambda}^{\text{sp}} = 2.17$ and $\tilde{\Lambda}^{\text{ch}} = 13.31$, are indicated by the arrows on the right.

In Fig.~\ref{fig: bilayer_2p}(b), we plot the bias dependent double occupancy fluctuation $\Delta D=\avg{({\hat{n}}_\uparrow - n_\uparrow ) ( {\hat{n}}_\downarrow - n_\downarrow )}$ (same for both layers), where $n_\uparrow n_\downarrow$ is subtracted from $\avg{\hat{n}_\uparrow \hat{n}_\downarrow}$.
The inset shows the layer dependent $\avg{\hat{n}_\uparrow \hat{n}_\downarrow}$, which is strongly influenced by the charge reshuffling.
As expected in a repulsively interacting system, $\Delta D<0$ due to the extra energy cost of double occupation.
In the equilibrium system with shifted local energies (dashed black line in panel (b)), the correlation effects decrease with increasing charge polarization, so that $\Delta D$ approaches zero with increasing $V$. 
In the nonequilibrium setup, a similar trend is evident for small $V$, and hence can be attributed to the displacement of the onsite energies.
As $V$ increases, the system traverses a complex nonequilibrium regime but for $V\gtrsim 2$,  $\Delta D$  starts to approach the single layer equilibrium value ($-0.114$), due to the effective decoupling of the layers at very large $V$. 

In Fig.~\ref{fig: bilayer_2p}(c) and (d), we plot the nearest-neighbor instaneous spin and charge fluctuations, respectively.
Also here, the term `fluctuation' refers to the fact that we substract the expectation values from the operators, i.e., plot $\avg{(\hat{X} - \avg{\hat{X}})(\hat{Y} - \avg{\hat{Y}})}$, where $X,Y \in \hat{N}_i, \hat{S}_i^z$.
This has no effect for the spin-spin correlations, since $\langle \hat{S}_i^z\rangle = 0$ in a paramagnetic state, but it shifts the charge-charge correlations by $N_i N_j$.
The red (blue) curves are for intra-layer (inter-layer) correlations.
The nonequilibrium results reveal that the intra-layer fluctuations (red curves), both in the spin and charge channels, are rather insensitive to $V$, while the inter-layer values (blue curves) react strongly to the voltage bias. 
Furthermore, the values and changes in the spin fluctuations are roughly three times larger than the equilibrium values and changes in the charge fluctuations, which is consistent with antiferromagnetically dominated short-range order in the structure. 
Remarkably, the inter-layer spin fluctuations (solid blue line in panel (c)) change sign around $V = 2.25$, indicating a switch of the preferred inter-layer spin alignment from antiferromagnetic (AFM) to ferromagnetic (FM).
The TPSC+GG results (not shown) exhibit the same switching from antiferromagnetic to ferromagnetic spin correlations with increasing $V$.
This switching cannot be explained by the electric field effect on the spin exchange coupling, which has been discussed in the case of Mott insulators \cite{Trotzky2008, Murakami2023}, since the sign inversion of the exchange interaction requires $V\approx U$.
Our observation is however consistent with Ref.~\cite{Dasari2019}, where the authors simulated the square lattice model subject to a short in-plane electric pulse using the FLEX+RPA method, and found a transient switch from antiferromagnetic to ferromagnetic correlations before relaxation. 

\begin{figure}
\includegraphics[width=1.0\linewidth]{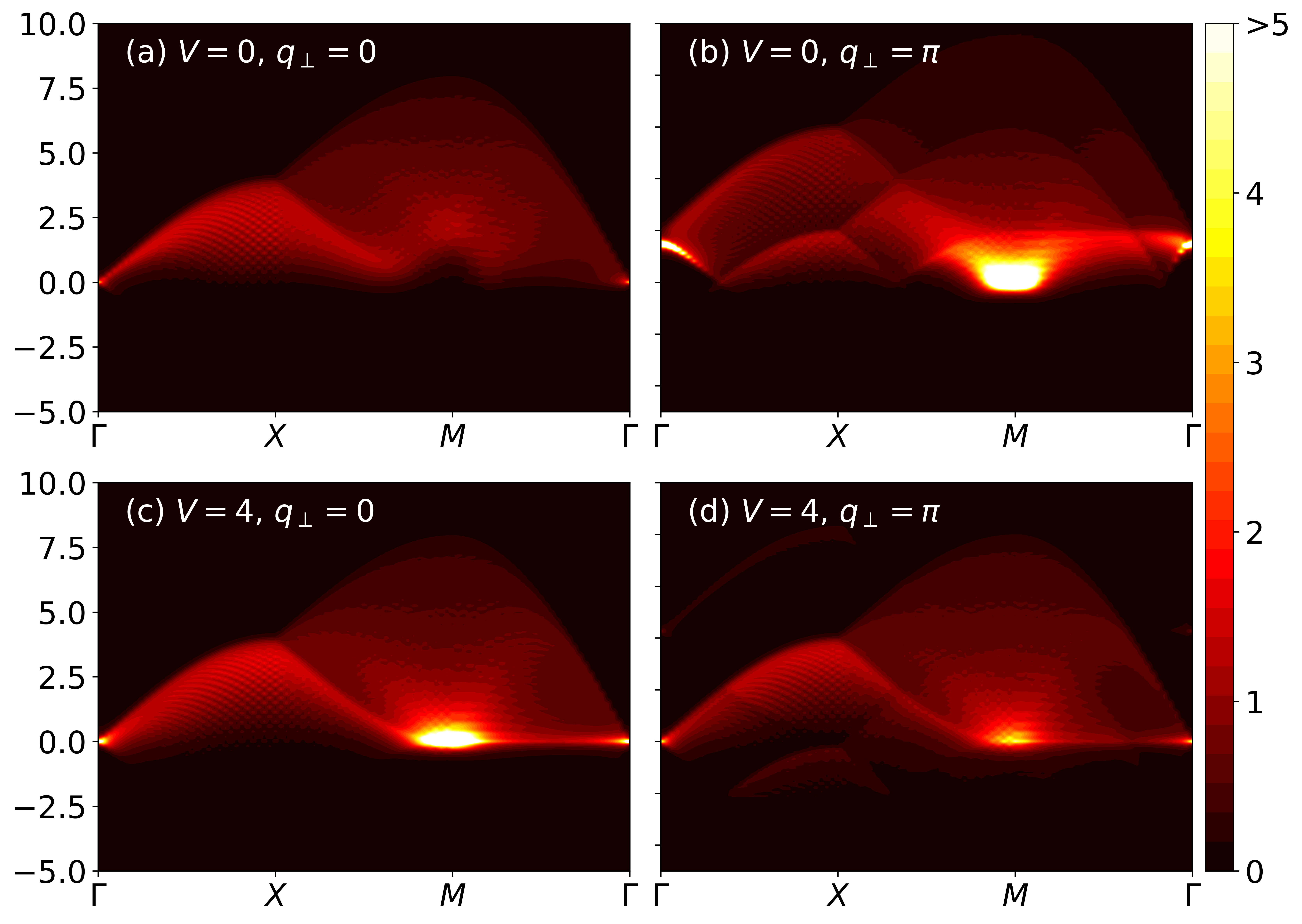}
\caption{Dynamical spin structure factor for $V=0$ (first row) and $V=4$ (second row), respectively.
The left column shows the symmetric ($q_\perp = 0$) channel and the right column the anti-symmetric ($q_\perp = \pi$) channel.
(Parameters: $W_1^{\perp}=1$, $U=4$, $\beta=4$.)
\label{fig: spin_strcut_factor}}
\end{figure}

To elucidate the spin dynamics, we perform calculations of the dynamical spin structure factor, also referred to as the (para)magnon dispersion.
This quantity is defined as the the Fourier transform of the non-local spin-spin correlation function, 
\begin{align}
&\sum_{\Tvec, T_\perp} \int \avg{\hat{S}_i^z(t) \hat{S}_{i+\Tvec +T_\perp}^z(t')} e^{i[ \qvec \cdot\Tvec + q_\perp T_\perp + \omega(t-t')]} d(t-t')\nonumber\\
&\hspace{20mm}= \sum_{T_\perp} e^{i q_\perp T_\perp} i\chi^{\text{sp},>}_{i,i+T_\perp}(\omega; \qvec)~,
\end{align}
where $T_\perp$ takes the value $0$ ($1$) for intra-layer (inter-layer) correlations.
The corresponding $q_\perp$ values are $0$ and $\pi$, which represent the symmetric (bonding) and anti-symmetric (antibonding) sectors.
In Fig.~\ref{fig: spin_strcut_factor}, we plot the dynamical spin structure factor for $V=0$ (first row) and $V=4$ (second row), respectively.
The left and right columns correspond to the symmetric ($q_\perp = 0$) and antisymmetric ($q_\perp = \pi$) channels.
In the equilibrium system, for $q_\perp=0$ (panel (a)), we see a linear dispersion around the $\Gamma$-point ($\qvec = (0,0)$), while the antisymmetric channel (panel (b)) exhibits a large signal near the $M$-point ($\qvec = (\pi,\pi)$).
These observations are qualitatively similar to the results for the bilayer Heisenberg model ($U \gg W$) obtained by QMC, see Fig.~1 in Ref.~\cite{Lohoefer2015}.
The most prominent effect of the bias $V=4$ (panels (c,d)) is a transfer of spectral weight from the anti-symmetric to the symmetric channel around the $M$-point, a result consistent 
with the observed switch from antiferromagnetic to ferromagnetic interlayer correlations. 
Given that the original TPSC method tends to overestimate correlation functions, it may be anticipated that these peaks will be smeared to some extent in more accurate calculations.

\begin{figure}
\includegraphics[width=1.0\linewidth]{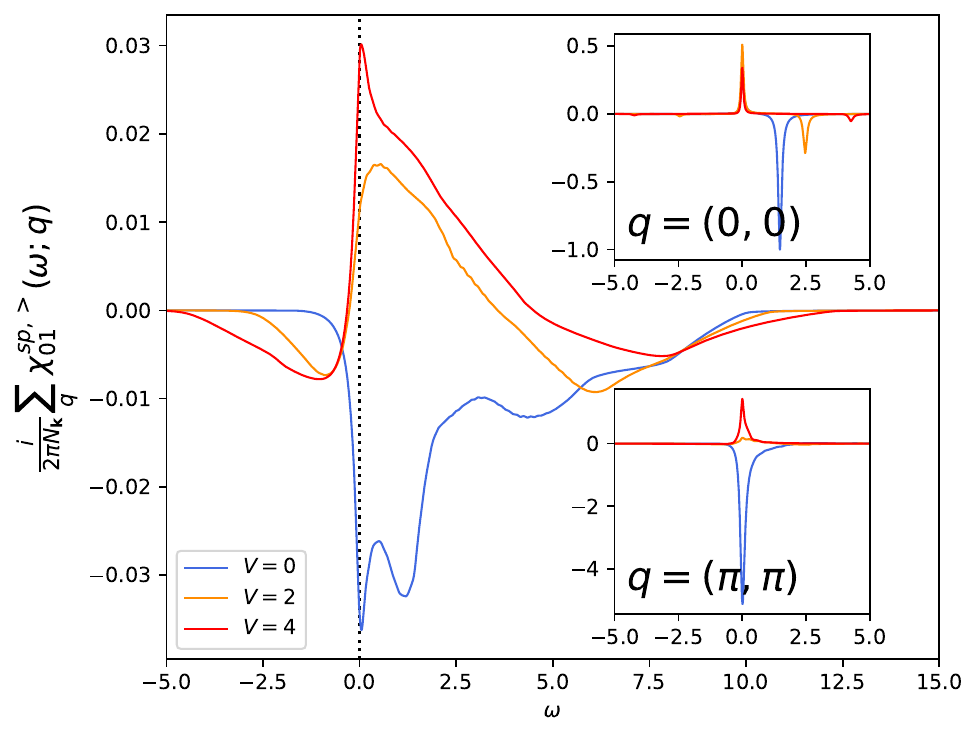}
\caption{Frequency-resolved and in-plane momentum averaged inter-layer spin susceptibility (greater component) for different voltage biases.
Insets: contributions from $\qvec = (0,0)$ and $\qvec = (\pi,\pi)$. (Parameters: $W_1^{\perp}=1$, $U=4$, $\beta=4$.)
\label{fig: w_resolved_chisp}}
\end{figure}

To gain further insights into the field-induced switching of the inter-layer spin correlations shown in Fig.~\ref{fig: bilayer_2p}(c), 
we plot in Fig.~\ref{fig: w_resolved_chisp} the momentum averaged inter-layer dynamical spin structure factor $\frac{i}{2\pi N_\kvec} \sum_{\qvec} \chi_{01}^{\text{sp},>}(\omega; \qvec)$.
This quantity represents the energy distribution of (in-plane momentum averaged) inter-layer spin excitations, and 
its integral over $\omega$ yields the instantaneous inter-layer spin correlation, see Eq.~\eqref{eq: real space spin-spin correlation formula}.
In equilibrium (blue curve), the spectrum is negative and dominated by two low-energy peaks.
The negative peak at $\omega = 0$ predominantly originates from momentum $\qvec = (\pi,\pi)$, and thus form a spin mode associated with both AFM inter-layer spin alignment in the bilayer structure (negative sign) and AFM intra-layer correlations ($\qvec$ vector), as can be seen from
\begin{dmath}
\avg{\hat{S}^z_{\Bvec}(t) \hat{S}^z_{\Bvec'+\Tvec}(t')} = \frac{1}{2\pi N_\kvec} \int d\omega \sum_\qvec i\chi^{\text{sp},>}_{\Bvec\Bvec'}(\omega; \qvec) e^{-i[\qvec\cdot\Tvec + \omega (t-t')]}~.
\end{dmath}
The above equation is a time non-local generalization of Eq.~\eqref{eq: real space spin-spin correlation formula}.
The other prominent negative peak at $\omega \approx 1.16$ can be linked to in-plane momentum $\qvec = (0,0)$, and thus to an intra-plane uniform but oscillating spin mode.
The net effect of these correlations is a strengthening of inter-layer singlet states within a unit cell.
In the presence of a perpendicular electric field $V=4$, the dominant mode at $\omega = 0$ and $\qvec = (\pi,\pi)$ changes its sign from negative to positive, thus contributing to an inter-layer FM spin alignment (while the associated intra-layer correlations remain AFM).
Moreover, the negative weight of the mode at $\qvec = (0,0)$ is suppressed in the presence of the voltage bias.
The result is a tendency to form inter-layer spin triplet states within the unit cells.

\begin{figure}
\includegraphics[width=1.0\linewidth]{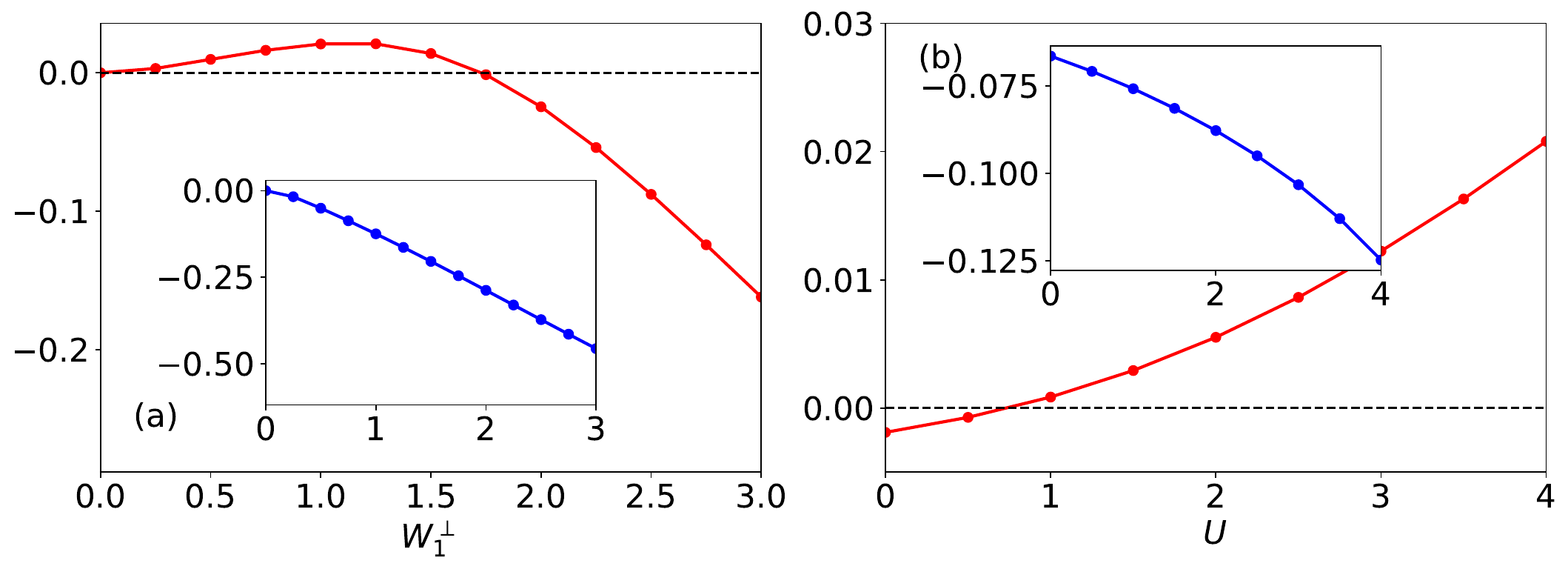}
\caption{
Inter-layer spin correlations for $V=4$ as a function of (a) inter-layer coupling $W_1^\perp$ (with $U=4$, $\beta=4$) and (b) Coulomb interaction $U$ (with $W_1^\perp=1$, $\beta=4$).
Insets: Corresponding equilibrium ($V=0$) results.
\label{fig: change tz}}
\end{figure}

Finally, we present in Fig.~\ref{fig: change tz} the inter-layer spin correlation $\avg{\hat{S}_0^z \hat{S}_1^z}$ as a function of inter-layer coupling $W_1^{\perp}$ (panel (a)) and as a function of the bare Coulomb interaction $U$ (panel (b)). 
The corresponding insets show the equilibrium results ($V=0$).
In panel (a), one can see that the interlayer spin correlation never crosses zero in equilibrium, and that the ferromagnetic interlayer correlations in the system with bias $V=4$ appear only in the hopping range $0 < W_1^\perp \lesssim 1.75 W_1^\parallel$.
For stronger $W_1^{\perp}$, antiferromagnetic interlayer correlations are recovered, since the electrons on different layers form a spin singlet.
The inter-layer spin correlations will approach $-0.5$ in the large-$W_1^{\perp}$ limit. (For a pure singlet state, we have $\avg{\hat{n}_{0\sigma} \hat{n}_{1\sigma}} = 0$ and hence $\avg{\hat{S}^z_{0} \hat{S}^z_{1}} = \avg{(\hat{n}_{0\uparrow} - \hat{n}_{0\downarrow})(\hat{n}_{1\uparrow} - \hat{n}_{1\downarrow})} = -2\avg{\hat{n}_{0\uparrow} \hat{n}_{1\downarrow}} = -0.5$.)

In panel (b), one can see that the inter-layer spin correlations monotonically increase with increasing $U$, undergoing a sign switch from negative to positive around $U=0.75$ for $V=4$.
In contrast, the equilibrium system shows increasing AFM correlations with increasing $U$.
Both observations show that the switch from antiferromagnetic to ferromagnetic spin correlations is a nontrivial correlated-electron phenomenon driven by the external electric field.

\section{Conclusions}\label{sec: conclusions}

We developed nonequilibrium steady-state two-particle self-consistent schemes, and used them to study spin and charge correlations in a bilayer square lattice Hubbard model subject to a perpendicular static electric field.
The TPSC method respects the Mermin-Wagner theorem, Pauli's exclusion principle and various sum rules by self-consistently calculating vertex functions which are approximated as local in both space and time.
From the local vertices, the non-local spatial correlations can then be obtained by solving the Schwinger-Dyson equation.
Our steady-state formalism is based on a two-branch Schwinger-Keldysh contour, and employs Fourier transformations to frequency-momentum space. It thus gives access to the spectral properties of the system both in and out of equilibrium without numerical analytical continuation.
In the applications, we mainly focused on the original TPSC method, which in contrast to TPSC+GG reproduces the pseudo-gap phenomenon related to short-ranged antiferromagnetic correlations.
For the equilibrium bilayer square lattice Hubbard model, we calculated the evolution of the spectral function with increasing inter-layer hopping, and illustrated the evolution from a system with antiferromagnetic pseudo-gap to a band insulator. 
We also demonstrated a good agreement of the calculated intra- and inter-layer correlations with data from recent optical lattice experiments. This shows that the TPSC approach can reliably predict the nonlocal correlations in the considered parameter regime. 

In the nonequilibrium calculations with perpendicular electric field, we observed that the system's behavior for small bias voltage is influenced by the charge reshuffling between the layers. With increasing bias or field strength, the system exhibits a non-trivial nonequilibrium behavior and finally approaches a state corresponding to decoupled equilibrium single layers in the limit of large $V$. The intra-layer spin and charge correlations are not significantly affected by the electric field, while the inter-layer correlations react strongly to the applied bias. 
Interestingly, we found that the inter-layer spin correlations switch from antiferromagnetic to ferromagnetic at some intermediate value of the perpendicular electric field when the intra- and inter-layer couplings are comparable. This switching is a nonequilibrium effect which can be linked to the inversion of the spectrum for collective spin excitations. 

\begin{acknowledgments}
The calculations were run on the Beo06 cluster at the University of Fribourg.
We acknowledge support from SNSF Grant No. 200021-196966 and thank O. Simard for helpful discussions. 
\end{acknowledgments}

\appendix

\section{Functional derivative technique\label{sec: functional derivative technique}}

We start from the partition function in the presence of an external source
\begin{dmath}
Z[\phi] = \text{Tr}\left[ T_\gamma\left( e^{-i\int_\gamma d\bar{z} \Hop(\bar{z})} \Sop[\phi] \right) \right]~,
\end{dmath}
where $T_\gamma$ is the time ordering operator on the contour $\gamma$, which orders the operators from right to left with increasing contour time (with additional minus sign for the exchange of two fermionic operators).
In the above equation, a generic non-diagonal source field term
\begin{dmath}
\Sop[\phi] = e^{-i\int_\gamma d\bar{z} d\bar{z}' \sum_\sigma \sum_{ij} c_{i\sigma}^\dag(\bar{z}) \phi_{ij,\sigma\sigma'}(\bar{z},\bar{z}') c_{j\sigma'}(\bar{z}')}~,
\end{dmath}
is introduced for the purpose of deriving thermodynamically consistent quantities.
Physical quantities can be obtained by setting $\phi = 0$ in the end.
To simplify the notation, we use a compressed notation $1 \equiv (i,z,\sigma)$, where $i$, $z$ and $\sigma$ are site, contour time and spin indices, respectively.
With this, the source field term can be rewritten as
\begin{dmath}
\Sop[\phi] = \exp\left\{-i\int d\bar{1} d\bar{2} c^{\dag}(\bar{1}) \phi(\bar{1},\bar{2}) c(\bar{2}) \right\}~.
\end{dmath}
 
The generating functional of the Green's function is defined as the logarithm of the partition function, $\mathcal{G}[\phi] = -\ln Z[\phi]$.
Its first-order derivative with respect to $\phi$ yields the single-particle Green's function, 
\begin{equation}
G(1,2;\phi) = -\frac{\delta \ln Z[\phi]}{\delta \phi(2,1)} = -i\avg{T_\gamma\{ c(1), c^\dag(2) \}}_{\phi}~,
\end{equation}
where we introduced the expectation value
\begin{dmath}
\avg{T_\gamma\{\cdots\}}_\phi = \frac{1}{Z[\phi]}\text{Tr}\left[ T_\gamma\left( e^{-i\int_\gamma d\bar{z} \Hop(\bar{z})} \Sop[\phi] \cdots \right) \right]~.
\end{dmath}
The derivative of $G(1,2;\phi)$ (corresponding to the second-order derivative of $\mathcal{G}[\phi]$) gives the two-particle exchange-correlation function
\begin{dmath}
L(1,2,3,4;\phi) = \frac{\delta G(1,3;\phi)}{\delta \phi(4,2^+)} = G(1,3;\phi) G(2,4;\phi) - G(1,2,3,4;\phi)~,
\end{dmath}
where
\begin{dmath}
G(1,2,3,4;\phi) = (-i)^2 \avg{T_\gamma\{ c(1) c(2) c^\dag(4) c^\dag(3) \}}_\phi~,
\end{dmath}
is the two-particle Green's function.
The generalized four-point susceptibility is then defined as
\begin{dmath}
\chi(1,2,3,4;\phi) = -iL(1,2,3,4;\phi)~.
\end{dmath}

While the Green's function can be generated by the functional derivative of $\mathcal{G}[\phi]$, one can also introduce a generating functional of the vertex, which is the Legendre transform of $\mathcal{G}[\phi]$, i.e.,
\begin{dmath}
\Gamma[G] = -\text{Tr}\left( G * \phi \right) - \ln Z[\phi]~.
\end{dmath}
Here, we introduced the trace operator $\text{Tr} A = \int d\bar{1} A(\bar{1},\bar{1})$ and the convolution $[A*B](1,2) = \int d\bar{3} A(1,\bar{3}) B(\bar{3},2)$.
The functional derivative of $\Gamma[G]$ with respect to $G$ gives the source field,
\begin{dmath}\label{eq: legendre transform derivative}
\frac{\delta \Gamma[G]}{\delta G(2,1)} = -\phi(1,2)~.
\end{dmath}
One usually separates the non-interacting contribution from $\Gamma[G]$, $\Gamma[G] = \Gamma^0[G] + \Phi[G]$, and thereby introduces the Luttinger-Ward functional $\Phi[G]$, which contains all the correlation contributions.
$\Gamma^0[G]$ can be explicitly computed since the non-interacting action is Gaussian.
Its explicit form reads $\Gamma^0[G] = -\text{Tr}\left[ (G^0)^{-1} * G - 1 \right] + \text{Tr} \ln (iG)$.
As a result, we have
\begin{dmath}\label{eq: Gamma functional}
\Gamma[G] = -\text{Tr}\left[ (G^0)^{-1} * G - 1 \right] + \text{Tr} \ln (iG) + \Phi[G]~.
\end{dmath}
$\Phi[G]$ is the generating functional of the two-particle irreducible vertices, and its first-order derivative gives the self-energy (one-particle irreducible vertex), 
\begin{dmath}
\Sigma(1,2) = \frac{\delta \Phi[G]}{\delta G(2,1)}~.
\end{dmath}
Applying the functional derivative with respect to $G$ to Eq.~\eqref{eq: Gamma functional}, one obtains the Dyson equation
\begin{dmath}\label{eq: Dyson equation with source field}
\frac{\delta \Gamma[G]}{\delta G(2,1^+)} = -[(G^0)^{-1}](1,2) + [G^{-1}](1,2) + \Sigma(1,2)~.
\end{dmath}
Note that the term on the left hand side is the source field, as seen from Eq.~\eqref{eq: legendre transform derivative}, and thus vanishes for the physical model.
The second-order derivative of $\Phi[G]$ with respect to $G$ yields the two-particle (particle-hole) irreducible vertex, 
\begin{equation}
\Lambda(1,2,3,4)
= -\frac{\delta^2 \Phi[G]}{\delta G(4,2) \delta G(3,1)}
= -\frac{\delta \Sigma(1,3)}{\delta G(4,2)}~.
\end{equation}
By taking the functional derivative of Eq.~\eqref{eq: Dyson equation with source field} and using the identities (variables with overbars are integrated over)
\begin{dmath}
\frac{\delta [G^{-1}](2,2')}{\delta G(1',1)} = - [G^{-1}](2,\bar{3})\frac{\delta G(\bar{3},\bar{4})}{\delta G(1',1)} [G^{-1}](\bar{4},2')~,
\end{dmath}
\begin{dmath}\label{eq: functional identity}
\frac{\delta^2 \ln Z[\phi]}{\delta \phi(\bar{6},\bar{5}) \delta \phi(2,1)} \frac{\delta^2 \Gamma[G]}{\delta G(4,3) \delta G(\bar{5},\bar{6})} = \delta(1-4) \delta(2-3)~,
\end{dmath}
we finally arrive at the Bethe-Salpeter equation for the generalized susceptibility,
\begin{dmath}\label{eq: Bethe-Salpeter for chi}
-i\chi(1,2,3,4) = -i G(1,4)G(2,3) + i G(1,\bar{1}) G(\bar{3},3) \Lambda(\bar{1},\bar{2},\bar{3},\bar{4}) \chi(\bar{4},2,\bar{2},4)~.
\end{dmath}
For an alternative derivation of these results, see Ref.~\cite{stefanucci2013nonequilibrium}.

\section{Consistency condition for $[\Sigma^{(2)} * G^{(1)}]$}\label{sec: tpsc consistency condition}

Here, we provide a brief proof of the nonequilibrium version of the consistency condition
\begin{dmath}\label{eq: nonequilibrium tpsc consistency condition}
[\Sigma^{(2)} * G^{(1)}](1,1^+) = iU(1)\avg{\hat{n}_\sigma^{(1)}(1) \hat{n}^{(1)}_{-\sigma}(1)}~,
\end{dmath}
where $\Sigma^{(2)}$ is calculated from Eq.~\eqref{eq: spectral self-energy} with $\alpha = 1$.

From the definition of $\Sigma^{(2)}$, we have
\begin{dmath}
[\Sigma^{(2)} * G^{(1)}](1,1^+) = iU(1) \left[ n(1) \right]^2 + i\frac{U(1)}{8} \frac{i}{2} \int d\bar{1} \chi^0(1,\bar{1})
\left[ \Lambda^{\text{ch}}(\bar{1}) \chi^{\text{ch}}(\bar{1},1) + 3 \Lambda^{\text{sp}}(\bar{1}) \chi^{\text{sp}}(\bar{1},1) \right]~,
\end{dmath}
where we have used $G(1,1^+) = i n^{(1)} (1)$ and Eq.~\eqref{eq: bare response function}.
With the relation $i\chi^0(1,1^+) = 2n^{(1)}(1) \left[ 1 - n^{(1)}(1) \right] $ and Eq.~\eqref{eq: simplified BSE}, after some algebra, one can prove Eq.~\eqref{eq: nonequilibrium tpsc consistency condition}.

\section{TPSC+GG single-layer spectrum}\label{sec: single layer tpsc+gg}

\begin{figure}
\includegraphics[width=1.0\linewidth]{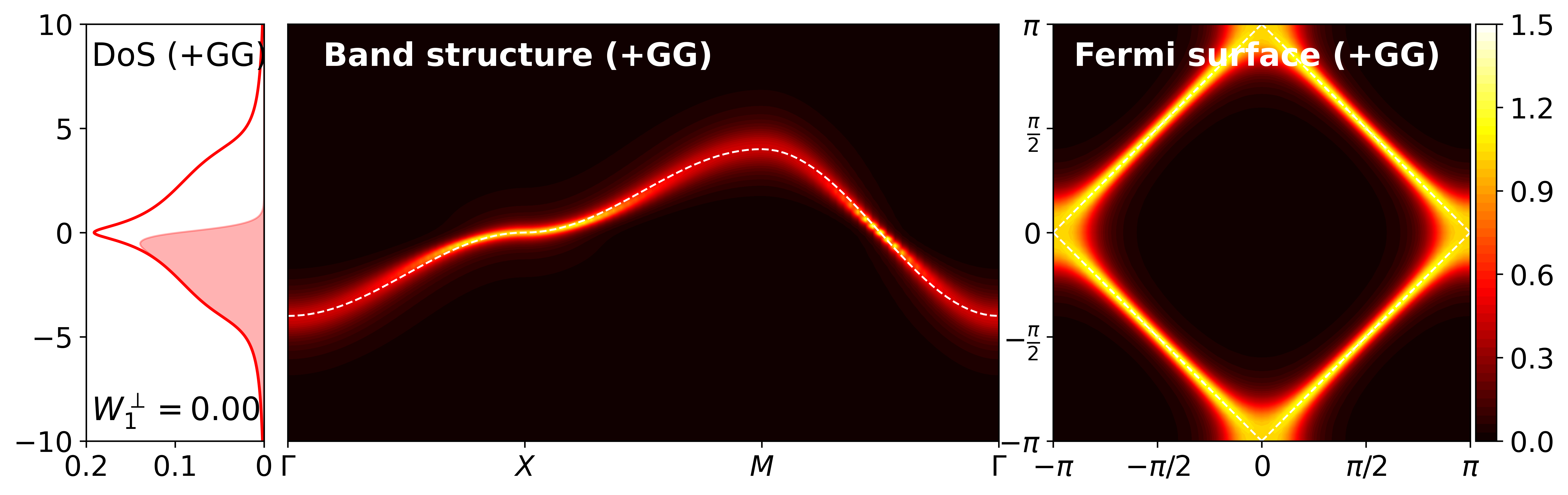 }
\caption{Spectral function, band structure and Fermi surface obtained by TPSC+GG for $W_1^{\perp} = 0$ (single layer) in equilibrium at $\beta=4$. 
These results can be compared to the corresponding TPSC data in Fig.~\ref{fig: tpsc_1Pqty_layer2_eq}(a).
\label{fig: tpscGG-1p}}
\end{figure}

Figure~\ref{fig: tpscGG-1p} plots the TPSC+GG results for the system with $W_1^{\perp} = 0$ in equilibrium, analogous to Fig.~\ref{fig: tpsc_1Pqty_layer2_eq}(a).
The local spectral function (left panel) and the momentum-resolved spectral function (middle panel) show no signs of an antiferromagnetic pseudo-gap, and also the Fermi surface estimated by $A({\bf k};\omega=0)$ (right panel) shows no clear suppression of the quasiparticles in the anti-nodal region. This indicates that as far as the pseudo-gap physics is concerned, TPSC+GG is less reliable in the present parameter regime than standard TPSC.
Inaccurate single-particle spectra are a common problem of self-consistent diagrammatic methods.
However, despite this limitation, recent studies have shown that TPSC+GG provides a significantly improved description of the two-particle correlation functions \cite{Schaefer2021,Simard2023}.

\begin{table}[t]
\begin{tabular}{cccccccc}
\hline
& $\tilde{\Lambda}^{\text{sp}}$ & $\tilde{\Lambda}^{\text{ch}}$ & $\avg{\hat{S}^z_0 \hat{S}^z_0}$ & $\avg{\hat{S}^z_0 \hat{S}^z_1}$ & $\avg{\hat{N}_0 \hat{N}_0}$ & $\avg{\hat{N}_0 \hat{N}_1}$ \\
\hline
NESS \hspace{2mm}&  2.245 \hspace{2mm}& 10.406 \hspace{2mm}& 0.719 \hspace{2mm}& -0.098 \hspace{2mm}& 0.281 \hspace{2mm}& -0.037 \\
EQ & 2.246 \hspace{2mm}& 10.299 \hspace{2mm}& 0.719 \hspace{2mm}& -0.097 \hspace{2mm}& 0.281 \hspace{2mm}& -0.036 \\
\hline
\end{tabular}
\caption{Comparison of vertices and correlation functions computed in the nonequilibrium steady state (NESS) for $V=1.0$ and in the equilibrium system with $T=T_{\text{eff}} = 0.685$ (EQ).}
\label{tab: heating}
\end{table}

\begin{figure}[b]
\includegraphics[width=1.0\linewidth]{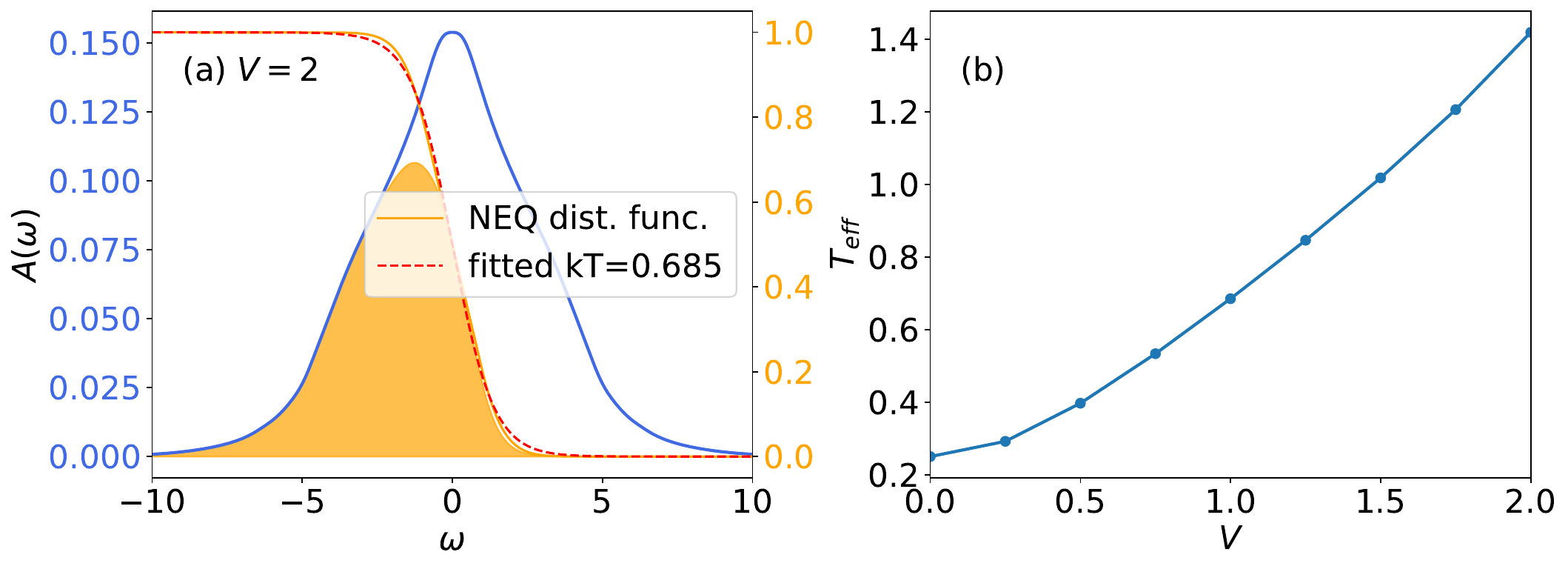}
\caption{Panel (a): local spectral function (in blue) and occupation (orange shaded area) for $V=1$. The orange solid line shows the nonequilibrium distribution function and the red dashed line a fitted Fermi function. Panel (b): effective temperature as a function of voltage bias $V$.}
\label{fig: layer1}
\end{figure}

\section{Bias induced heating in a single-layer lattice}\label{sec: single layer heating}

In this appendix, we study a single-layer square lattice under a perpendicular static electric field.
The parameters are the same as in the bilayer case ($U = 4$, $\beta = 4$ and $\Gamma = 0.05$).
In panel (a) of Fig.~\ref{fig: layer1}, we plot the spectral function for $V=1$ (blue), the occupation (orange shading), as well as the corresponding nonequilibrium distribution function ${f}_{\text{neq}}= - \Im G^< / [ 2\Im G^r ]$ (orange line).
By fitting ${f}_{\text{neq}}(\omega)$ with a Fermi function (red dashed line), one obtains the effective temperature $T_{{\rm eff}} = 0.685$ of the nonequilibrium state.
The good agreement between ${f}_{\text{neq}}(\omega)$ and the Fermi function indicates that the effect of a small bias is essentially a heating of the single-layer model ($T_{\rm eq}=0.25$).
In panel (b), we plot the effective temperature as a function of voltage bias.
Up to $V=2$, the effective temperature description can capture the NESS characteristics of the system.

To further validate the effective temperature description, we compared the vertices as well as local and nearest-neighbor correlations.
In Table~\ref{tab: heating}, the first row shows the result for the nonequilibrium system with bias $V=1$ and the second row the results for the equilibrium system with $T_{\text{eff}} = 0.685$. Again, one finds a remarkable agreement between the two systems.

\bibliography{ref}

\begin{thebibliography}{63}%
\makeatletter
\providecommand \@ifxundefined [1]{%
 \@ifx{#1\undefined}
}%
\providecommand \@ifnum [1]{%
 \ifnum #1\expandafter \@firstoftwo
 \else \expandafter \@secondoftwo
 \fi
}%
\providecommand \@ifx [1]{%
 \ifx #1\expandafter \@firstoftwo
 \else \expandafter \@secondoftwo
 \fi
}%
\providecommand \natexlab [1]{#1}%
\providecommand \enquote  [1]{``#1''}%
\providecommand \bibnamefont  [1]{#1}%
\providecommand \bibfnamefont [1]{#1}%
\providecommand \citenamefont [1]{#1}%
\providecommand \href@noop [0]{\@secondoftwo}%
\providecommand \href [0]{\begingroup \@sanitize@url \@href}%
\providecommand \@href[1]{\@@startlink{#1}\@@href}%
\providecommand \@@href[1]{\endgroup#1\@@endlink}%
\providecommand \@sanitize@url [0]{\catcode `\\12\catcode `\$12\catcode
  `\&12\catcode `\#12\catcode `\^12\catcode `\_12\catcode `\%12\relax}%
\providecommand \@@startlink[1]{}%
\providecommand \@@endlink[0]{}%
\providecommand \url  [0]{\begingroup\@sanitize@url \@url }%
\providecommand \@url [1]{\endgroup\@href {#1}{\urlprefix }}%
\providecommand \urlprefix  [0]{URL }%
\providecommand \Eprint [0]{\href }%
\providecommand \doibase [0]{https://doi.org/}%
\providecommand \selectlanguage [0]{\@gobble}%
\providecommand \bibinfo  [0]{\@secondoftwo}%
\providecommand \bibfield  [0]{\@secondoftwo}%
\providecommand \translation [1]{[#1]}%
\providecommand \BibitemOpen [0]{}%
\providecommand \bibitemStop [0]{}%
\providecommand \bibitemNoStop [0]{.\EOS\space}%
\providecommand \EOS [0]{\spacefactor3000\relax}%
\providecommand \BibitemShut  [1]{\csname bibitem#1\endcsname}%
\let\auto@bib@innerbib\@empty
\bibitem [{\citenamefont {Keimer}\ \emph {et~al.}(2015)\citenamefont {Keimer},
  \citenamefont {Kivelson}, \citenamefont {Norman}, \citenamefont {Uchida},\
  and\ \citenamefont {Zaanen}}]{Keimer2015}%
  \BibitemOpen
  \bibfield  {author} {\bibinfo {author} {\bibfnamefont {B.}~\bibnamefont
  {Keimer}}, \bibinfo {author} {\bibfnamefont {S.~A.}\ \bibnamefont
  {Kivelson}}, \bibinfo {author} {\bibfnamefont {M.~R.}\ \bibnamefont
  {Norman}}, \bibinfo {author} {\bibfnamefont {S.}~\bibnamefont {Uchida}},\
  and\ \bibinfo {author} {\bibfnamefont {J.}~\bibnamefont {Zaanen}},\ }\href
  {https://doi.org/10.1038/nature14165} {\bibfield  {journal} {\bibinfo
  {journal} {Nature}\ }\textbf {\bibinfo {volume} {518}},\ \bibinfo {pages}
  {179} (\bibinfo {year} {2015})}\BibitemShut {NoStop}%
\bibitem [{\citenamefont {Galanakis}\ \emph {et~al.}(2011)\citenamefont
  {Galanakis}, \citenamefont {Khatami}, \citenamefont {Mikelsons},
  \citenamefont {Macridin}, \citenamefont {Moreno}, \citenamefont {Browne},\
  and\ \citenamefont {Jarrell}}]{Galanakis2011}%
  \BibitemOpen
  \bibfield  {author} {\bibinfo {author} {\bibfnamefont {D.}~\bibnamefont
  {Galanakis}}, \bibinfo {author} {\bibfnamefont {E.}~\bibnamefont {Khatami}},
  \bibinfo {author} {\bibfnamefont {K.}~\bibnamefont {Mikelsons}}, \bibinfo
  {author} {\bibfnamefont {A.}~\bibnamefont {Macridin}}, \bibinfo {author}
  {\bibfnamefont {J.}~\bibnamefont {Moreno}}, \bibinfo {author} {\bibfnamefont
  {D.~A.}\ \bibnamefont {Browne}},\ and\ \bibinfo {author} {\bibfnamefont
  {M.}~\bibnamefont {Jarrell}},\ }\href
  {https://doi.org/10.1098/rsta.2010.0228} {\bibfield  {journal} {\bibinfo
  {journal} {Philosophical Transactions of the Royal Society A: Mathematical,
  Physical and Engineering Sciences}\ }\textbf {\bibinfo {volume} {369}},\
  \bibinfo {pages} {1670} (\bibinfo {year} {2011})}\BibitemShut {NoStop}%
\bibitem [{\citenamefont {Aoki}\ \emph {et~al.}(2014)\citenamefont {Aoki},
  \citenamefont {Tsuji}, \citenamefont {Eckstein}, \citenamefont {Kollar},
  \citenamefont {Oka},\ and\ \citenamefont {Werner}}]{Aoki2014}%
  \BibitemOpen
  \bibfield  {author} {\bibinfo {author} {\bibfnamefont {H.}~\bibnamefont
  {Aoki}}, \bibinfo {author} {\bibfnamefont {N.}~\bibnamefont {Tsuji}},
  \bibinfo {author} {\bibfnamefont {M.}~\bibnamefont {Eckstein}}, \bibinfo
  {author} {\bibfnamefont {M.}~\bibnamefont {Kollar}}, \bibinfo {author}
  {\bibfnamefont {T.}~\bibnamefont {Oka}},\ and\ \bibinfo {author}
  {\bibfnamefont {P.}~\bibnamefont {Werner}},\ }\href
  {https://doi.org/10.1103/RevModPhys.86.779} {\bibfield  {journal} {\bibinfo
  {journal} {Rev. Mod. Phys.}\ }\textbf {\bibinfo {volume} {86}},\ \bibinfo
  {pages} {779} (\bibinfo {year} {2014})}\BibitemShut {NoStop}%
\bibitem [{\citenamefont {Murakami}\ \emph {et~al.}(2023)\citenamefont
  {Murakami}, \citenamefont {Golež}, \citenamefont {Eckstein},\ and\
  \citenamefont {Werner}}]{Murakami2023}%
  \BibitemOpen
  \bibfield  {author} {\bibinfo {author} {\bibfnamefont {Y.}~\bibnamefont
  {Murakami}}, \bibinfo {author} {\bibfnamefont {D.}~\bibnamefont {Golež}},
  \bibinfo {author} {\bibfnamefont {M.}~\bibnamefont {Eckstein}},\ and\
  \bibinfo {author} {\bibfnamefont {P.}~\bibnamefont {Werner}},\ }\href@noop {}
  {\  (\bibinfo {year} {2023})},\ \Eprint {https://arxiv.org/abs/2310.05201}
  {arXiv:2310.05201 [cond-mat.str-el]} \BibitemShut {NoStop}%
\bibitem [{\citenamefont {LeBlanc}\ \emph {et~al.}(2015)\citenamefont
  {LeBlanc}, \citenamefont {Antipov}, \citenamefont {Becca}, \citenamefont
  {Bulik}, \citenamefont {Chan}, \citenamefont {Chung}, \citenamefont {Deng},
  \citenamefont {Ferrero}, \citenamefont {Henderson}, \citenamefont
  {Jim\'enez-Hoyos}, \citenamefont {Kozik}, \citenamefont {Liu}, \citenamefont
  {Millis}, \citenamefont {Prokof'ev}, \citenamefont {Qin}, \citenamefont
  {Scuseria}, \citenamefont {Shi}, \citenamefont {Svistunov}, \citenamefont
  {Tocchio}, \citenamefont {Tupitsyn}, \citenamefont {White}, \citenamefont
  {Zhang}, \citenamefont {Zheng}, \citenamefont {Zhu},\ and\ \citenamefont
  {Gull}}]{LeBlanc2015}%
  \BibitemOpen
  \bibfield  {author} {\bibinfo {author} {\bibfnamefont {J.~P.~F.}\
  \bibnamefont {LeBlanc}}, \bibinfo {author} {\bibfnamefont {A.~E.}\
  \bibnamefont {Antipov}}, \bibinfo {author} {\bibfnamefont {F.}~\bibnamefont
  {Becca}}, \bibinfo {author} {\bibfnamefont {I.~W.}\ \bibnamefont {Bulik}},
  \bibinfo {author} {\bibfnamefont {G.~K.-L.}\ \bibnamefont {Chan}}, \bibinfo
  {author} {\bibfnamefont {C.-M.}\ \bibnamefont {Chung}}, \bibinfo {author}
  {\bibfnamefont {Y.}~\bibnamefont {Deng}}, \bibinfo {author} {\bibfnamefont
  {M.}~\bibnamefont {Ferrero}}, \bibinfo {author} {\bibfnamefont {T.~M.}\
  \bibnamefont {Henderson}}, \bibinfo {author} {\bibfnamefont {C.~A.}\
  \bibnamefont {Jim\'enez-Hoyos}}, \bibinfo {author} {\bibfnamefont
  {E.}~\bibnamefont {Kozik}}, \bibinfo {author} {\bibfnamefont {X.-W.}\
  \bibnamefont {Liu}}, \bibinfo {author} {\bibfnamefont {A.~J.}\ \bibnamefont
  {Millis}}, \bibinfo {author} {\bibfnamefont {N.~V.}\ \bibnamefont
  {Prokof'ev}}, \bibinfo {author} {\bibfnamefont {M.}~\bibnamefont {Qin}},
  \bibinfo {author} {\bibfnamefont {G.~E.}\ \bibnamefont {Scuseria}}, \bibinfo
  {author} {\bibfnamefont {H.}~\bibnamefont {Shi}}, \bibinfo {author}
  {\bibfnamefont {B.~V.}\ \bibnamefont {Svistunov}}, \bibinfo {author}
  {\bibfnamefont {L.~F.}\ \bibnamefont {Tocchio}}, \bibinfo {author}
  {\bibfnamefont {I.~S.}\ \bibnamefont {Tupitsyn}}, \bibinfo {author}
  {\bibfnamefont {S.~R.}\ \bibnamefont {White}}, \bibinfo {author}
  {\bibfnamefont {S.}~\bibnamefont {Zhang}}, \bibinfo {author} {\bibfnamefont
  {B.-X.}\ \bibnamefont {Zheng}}, \bibinfo {author} {\bibfnamefont
  {Z.}~\bibnamefont {Zhu}},\ and\ \bibinfo {author} {\bibfnamefont
  {E.}~\bibnamefont {Gull}} (\bibinfo {collaboration} {Simons Collaboration on
  the Many-Electron Problem}),\ }\href
  {https://doi.org/10.1103/PhysRevX.5.041041} {\bibfield  {journal} {\bibinfo
  {journal} {Phys. Rev. X}\ }\textbf {\bibinfo {volume} {5}},\ \bibinfo {pages}
  {041041} (\bibinfo {year} {2015})}\BibitemShut {NoStop}%
\bibitem [{\citenamefont {Qin}\ \emph {et~al.}(2022)\citenamefont {Qin},
  \citenamefont {Sch\"{a}fer}, \citenamefont {Andergassen}, \citenamefont
  {Corboz},\ and\ \citenamefont {Gull}}]{Qin2022}%
  \BibitemOpen
  \bibfield  {author} {\bibinfo {author} {\bibfnamefont {M.}~\bibnamefont
  {Qin}}, \bibinfo {author} {\bibfnamefont {T.}~\bibnamefont {Sch\"{a}fer}},
  \bibinfo {author} {\bibfnamefont {S.}~\bibnamefont {Andergassen}}, \bibinfo
  {author} {\bibfnamefont {P.}~\bibnamefont {Corboz}},\ and\ \bibinfo {author}
  {\bibfnamefont {E.}~\bibnamefont {Gull}},\ }\href
  {https://doi.org/10.1146/annurev-conmatphys-090921-033948} {\bibfield
  {journal} {\bibinfo  {journal} {Annual Review of Condensed Matter Physics}\
  }\textbf {\bibinfo {volume} {13}},\ \bibinfo {pages} {275} (\bibinfo {year}
  {2022})}\BibitemShut {NoStop}%
\bibitem [{\citenamefont {Sch\"afer}\ \emph {et~al.}(2021)\citenamefont
  {Sch\"afer}, \citenamefont {Wentzell}, \citenamefont {\ifmmode~\check{S}\else
  \v{S}\fi{}imkovic}, \citenamefont {He}, \citenamefont {Hille}, \citenamefont
  {Klett}, \citenamefont {Eckhardt}, \citenamefont {Arzhang}, \citenamefont
  {Harkov}, \citenamefont {Le~R\'egent}, \citenamefont {Kirsch}, \citenamefont
  {Wang}, \citenamefont {Kim}, \citenamefont {Kozik}, \citenamefont {Stepanov},
  \citenamefont {Kauch}, \citenamefont {Andergassen}, \citenamefont {Hansmann},
  \citenamefont {Rohe}, \citenamefont {Vilk}, \citenamefont {LeBlanc},
  \citenamefont {Zhang}, \citenamefont {Tremblay}, \citenamefont {Ferrero},
  \citenamefont {Parcollet},\ and\ \citenamefont {Georges}}]{Schaefer2021}%
  \BibitemOpen
  \bibfield  {author} {\bibinfo {author} {\bibfnamefont {T.}~\bibnamefont
  {Sch\"afer}}, \bibinfo {author} {\bibfnamefont {N.}~\bibnamefont {Wentzell}},
  \bibinfo {author} {\bibfnamefont {F.}~\bibnamefont {\ifmmode~\check{S}\else
  \v{S}\fi{}imkovic}}, \bibinfo {author} {\bibfnamefont {Y.-Y.}\ \bibnamefont
  {He}}, \bibinfo {author} {\bibfnamefont {C.}~\bibnamefont {Hille}}, \bibinfo
  {author} {\bibfnamefont {M.}~\bibnamefont {Klett}}, \bibinfo {author}
  {\bibfnamefont {C.~J.}\ \bibnamefont {Eckhardt}}, \bibinfo {author}
  {\bibfnamefont {B.}~\bibnamefont {Arzhang}}, \bibinfo {author} {\bibfnamefont
  {V.}~\bibnamefont {Harkov}}, \bibinfo {author} {\bibfnamefont {F.~m. c.-M.}\
  \bibnamefont {Le~R\'egent}}, \bibinfo {author} {\bibfnamefont
  {A.}~\bibnamefont {Kirsch}}, \bibinfo {author} {\bibfnamefont
  {Y.}~\bibnamefont {Wang}}, \bibinfo {author} {\bibfnamefont {A.~J.}\
  \bibnamefont {Kim}}, \bibinfo {author} {\bibfnamefont {E.}~\bibnamefont
  {Kozik}}, \bibinfo {author} {\bibfnamefont {E.~A.}\ \bibnamefont {Stepanov}},
  \bibinfo {author} {\bibfnamefont {A.}~\bibnamefont {Kauch}}, \bibinfo
  {author} {\bibfnamefont {S.}~\bibnamefont {Andergassen}}, \bibinfo {author}
  {\bibfnamefont {P.}~\bibnamefont {Hansmann}}, \bibinfo {author}
  {\bibfnamefont {D.}~\bibnamefont {Rohe}}, \bibinfo {author} {\bibfnamefont
  {Y.~M.}\ \bibnamefont {Vilk}}, \bibinfo {author} {\bibfnamefont {J.~P.~F.}\
  \bibnamefont {LeBlanc}}, \bibinfo {author} {\bibfnamefont {S.}~\bibnamefont
  {Zhang}}, \bibinfo {author} {\bibfnamefont {A.-M.~S.}\ \bibnamefont
  {Tremblay}}, \bibinfo {author} {\bibfnamefont {M.}~\bibnamefont {Ferrero}},
  \bibinfo {author} {\bibfnamefont {O.}~\bibnamefont {Parcollet}},\ and\
  \bibinfo {author} {\bibfnamefont {A.}~\bibnamefont {Georges}},\ }\href
  {https://doi.org/10.1103/PhysRevX.11.011058} {\bibfield  {journal} {\bibinfo
  {journal} {Phys. Rev. X}\ }\textbf {\bibinfo {volume} {11}},\ \bibinfo
  {pages} {011058} (\bibinfo {year} {2021})}\BibitemShut {NoStop}%
\bibitem [{\citenamefont {Or{\'u}s}(2019)}]{Orus2019}%
  \BibitemOpen
  \bibfield  {author} {\bibinfo {author} {\bibfnamefont {R.}~\bibnamefont
  {Or{\'u}s}},\ }\href {https://doi.org/10.1038/s42254-019-0086-7} {\bibfield
  {journal} {\bibinfo  {journal} {Nature Reviews Physics}\ }\textbf {\bibinfo
  {volume} {1}},\ \bibinfo {pages} {538} (\bibinfo {year} {2019})}\BibitemShut
  {NoStop}%
\bibitem [{\citenamefont {Metzner}\ and\ \citenamefont
  {Vollhardt}(1989)}]{Metzner1989}%
  \BibitemOpen
  \bibfield  {author} {\bibinfo {author} {\bibfnamefont {W.}~\bibnamefont
  {Metzner}}\ and\ \bibinfo {author} {\bibfnamefont {D.}~\bibnamefont
  {Vollhardt}},\ }\href {https://doi.org/10.1103/PhysRevLett.62.324} {\bibfield
   {journal} {\bibinfo  {journal} {Phys. Rev. Lett.}\ }\textbf {\bibinfo
  {volume} {62}},\ \bibinfo {pages} {324} (\bibinfo {year} {1989})}\BibitemShut
  {NoStop}%
\bibitem [{\citenamefont {Georges}\ and\ \citenamefont
  {Kotliar}(1992)}]{Georges1992}%
  \BibitemOpen
  \bibfield  {author} {\bibinfo {author} {\bibfnamefont {A.}~\bibnamefont
  {Georges}}\ and\ \bibinfo {author} {\bibfnamefont {G.}~\bibnamefont
  {Kotliar}},\ }\href {https://doi.org/10.1103/PhysRevB.45.6479} {\bibfield
  {journal} {\bibinfo  {journal} {Phys. Rev. B}\ }\textbf {\bibinfo {volume}
  {45}},\ \bibinfo {pages} {6479} (\bibinfo {year} {1992})}\BibitemShut
  {NoStop}%
\bibitem [{\citenamefont {Georges}\ \emph {et~al.}(1996)\citenamefont
  {Georges}, \citenamefont {Kotliar}, \citenamefont {Krauth},\ and\
  \citenamefont {Rozenberg}}]{Georges1996}%
  \BibitemOpen
  \bibfield  {author} {\bibinfo {author} {\bibfnamefont {A.}~\bibnamefont
  {Georges}}, \bibinfo {author} {\bibfnamefont {G.}~\bibnamefont {Kotliar}},
  \bibinfo {author} {\bibfnamefont {W.}~\bibnamefont {Krauth}},\ and\ \bibinfo
  {author} {\bibfnamefont {M.~J.}\ \bibnamefont {Rozenberg}},\ }\href
  {https://doi.org/10.1103/RevModPhys.68.13} {\bibfield  {journal} {\bibinfo
  {journal} {Rev. Mod. Phys.}\ }\textbf {\bibinfo {volume} {68}},\ \bibinfo
  {pages} {13} (\bibinfo {year} {1996})}\BibitemShut {NoStop}%
\bibitem [{\citenamefont {Kotliar}\ and\ \citenamefont
  {Vollhardt}(2004)}]{Kotliar2004}%
  \BibitemOpen
  \bibfield  {author} {\bibinfo {author} {\bibfnamefont {G.}~\bibnamefont
  {Kotliar}}\ and\ \bibinfo {author} {\bibfnamefont {D.}~\bibnamefont
  {Vollhardt}},\ }\href {https://doi.org/10.1063/1.1712502} {\bibfield
  {journal} {\bibinfo  {journal} {Physics Today}\ }\textbf {\bibinfo {volume}
  {57}},\ \bibinfo {pages} {53} (\bibinfo {year} {2004})}\BibitemShut {NoStop}%
\bibitem [{\citenamefont {Gull}\ \emph {et~al.}(2011)\citenamefont {Gull},
  \citenamefont {Millis}, \citenamefont {Lichtenstein}, \citenamefont
  {Rubtsov}, \citenamefont {Troyer},\ and\ \citenamefont {Werner}}]{Gull2011}%
  \BibitemOpen
  \bibfield  {author} {\bibinfo {author} {\bibfnamefont {E.}~\bibnamefont
  {Gull}}, \bibinfo {author} {\bibfnamefont {A.~J.}\ \bibnamefont {Millis}},
  \bibinfo {author} {\bibfnamefont {A.~I.}\ \bibnamefont {Lichtenstein}},
  \bibinfo {author} {\bibfnamefont {A.~N.}\ \bibnamefont {Rubtsov}}, \bibinfo
  {author} {\bibfnamefont {M.}~\bibnamefont {Troyer}},\ and\ \bibinfo {author}
  {\bibfnamefont {P.}~\bibnamefont {Werner}},\ }\href
  {https://doi.org/10.1103/RevModPhys.83.349} {\bibfield  {journal} {\bibinfo
  {journal} {Rev. Mod. Phys.}\ }\textbf {\bibinfo {volume} {83}},\ \bibinfo
  {pages} {349} (\bibinfo {year} {2011})}\BibitemShut {NoStop}%
\bibitem [{\citenamefont {Werner}\ \emph {et~al.}(2006)\citenamefont {Werner},
  \citenamefont {Comanac}, \citenamefont {de' Medici}, \citenamefont {Troyer},\
  and\ \citenamefont {Millis}}]{Werner2006}%
  \BibitemOpen
  \bibfield  {author} {\bibinfo {author} {\bibfnamefont {P.}~\bibnamefont
  {Werner}}, \bibinfo {author} {\bibfnamefont {A.}~\bibnamefont {Comanac}},
  \bibinfo {author} {\bibfnamefont {L.}~\bibnamefont {de' Medici}}, \bibinfo
  {author} {\bibfnamefont {M.}~\bibnamefont {Troyer}},\ and\ \bibinfo {author}
  {\bibfnamefont {A.~J.}\ \bibnamefont {Millis}},\ }\href
  {https://doi.org/10.1103/PhysRevLett.97.076405} {\bibfield  {journal}
  {\bibinfo  {journal} {Phys. Rev. Lett.}\ }\textbf {\bibinfo {volume} {97}},\
  \bibinfo {pages} {076405} (\bibinfo {year} {2006})}\BibitemShut {NoStop}%
\bibitem [{\citenamefont {Wilson}(1975)}]{Wilson1975}%
  \BibitemOpen
  \bibfield  {author} {\bibinfo {author} {\bibfnamefont {K.~G.}\ \bibnamefont
  {Wilson}},\ }\href {https://doi.org/10.1103/RevModPhys.47.773} {\bibfield
  {journal} {\bibinfo  {journal} {Rev. Mod. Phys.}\ }\textbf {\bibinfo {volume}
  {47}},\ \bibinfo {pages} {773} (\bibinfo {year} {1975})}\BibitemShut
  {NoStop}%
\bibitem [{\citenamefont {Bulla}\ \emph {et~al.}(2008)\citenamefont {Bulla},
  \citenamefont {Costi},\ and\ \citenamefont {Pruschke}}]{Bulla2008}%
  \BibitemOpen
  \bibfield  {author} {\bibinfo {author} {\bibfnamefont {R.}~\bibnamefont
  {Bulla}}, \bibinfo {author} {\bibfnamefont {T.~A.}\ \bibnamefont {Costi}},\
  and\ \bibinfo {author} {\bibfnamefont {T.}~\bibnamefont {Pruschke}},\ }\href
  {https://doi.org/10.1103/RevModPhys.80.395} {\bibfield  {journal} {\bibinfo
  {journal} {Rev. Mod. Phys.}\ }\textbf {\bibinfo {volume} {80}},\ \bibinfo
  {pages} {395} (\bibinfo {year} {2008})}\BibitemShut {NoStop}%
\bibitem [{\citenamefont {Maier}\ \emph {et~al.}(2005)\citenamefont {Maier},
  \citenamefont {Jarrell}, \citenamefont {Pruschke},\ and\ \citenamefont
  {Hettler}}]{Maier2005}%
  \BibitemOpen
  \bibfield  {author} {\bibinfo {author} {\bibfnamefont {T.}~\bibnamefont
  {Maier}}, \bibinfo {author} {\bibfnamefont {M.}~\bibnamefont {Jarrell}},
  \bibinfo {author} {\bibfnamefont {T.}~\bibnamefont {Pruschke}},\ and\
  \bibinfo {author} {\bibfnamefont {M.~H.}\ \bibnamefont {Hettler}},\ }\href
  {https://doi.org/10.1103/RevModPhys.77.1027} {\bibfield  {journal} {\bibinfo
  {journal} {Rev. Mod. Phys.}\ }\textbf {\bibinfo {volume} {77}},\ \bibinfo
  {pages} {1027} (\bibinfo {year} {2005})}\BibitemShut {NoStop}%
\bibitem [{\citenamefont {Rohringer}\ \emph {et~al.}(2018)\citenamefont
  {Rohringer}, \citenamefont {Hafermann}, \citenamefont {Toschi}, \citenamefont
  {Katanin}, \citenamefont {Antipov}, \citenamefont {Katsnelson}, \citenamefont
  {Lichtenstein}, \citenamefont {Rubtsov},\ and\ \citenamefont
  {Held}}]{Rohringer2018}%
  \BibitemOpen
  \bibfield  {author} {\bibinfo {author} {\bibfnamefont {G.}~\bibnamefont
  {Rohringer}}, \bibinfo {author} {\bibfnamefont {H.}~\bibnamefont
  {Hafermann}}, \bibinfo {author} {\bibfnamefont {A.}~\bibnamefont {Toschi}},
  \bibinfo {author} {\bibfnamefont {A.~A.}\ \bibnamefont {Katanin}}, \bibinfo
  {author} {\bibfnamefont {A.~E.}\ \bibnamefont {Antipov}}, \bibinfo {author}
  {\bibfnamefont {M.~I.}\ \bibnamefont {Katsnelson}}, \bibinfo {author}
  {\bibfnamefont {A.~I.}\ \bibnamefont {Lichtenstein}}, \bibinfo {author}
  {\bibfnamefont {A.~N.}\ \bibnamefont {Rubtsov}},\ and\ \bibinfo {author}
  {\bibfnamefont {K.}~\bibnamefont {Held}},\ }\href
  {https://doi.org/10.1103/RevModPhys.90.025003} {\bibfield  {journal}
  {\bibinfo  {journal} {Rev. Mod. Phys.}\ }\textbf {\bibinfo {volume} {90}},\
  \bibinfo {pages} {025003} (\bibinfo {year} {2018})}\BibitemShut {NoStop}%
\bibitem [{\citenamefont {Tsuji}\ \emph {et~al.}(2014)\citenamefont {Tsuji},
  \citenamefont {Barmettler}, \citenamefont {Aoki},\ and\ \citenamefont
  {Werner}}]{Tsuji2014}%
  \BibitemOpen
  \bibfield  {author} {\bibinfo {author} {\bibfnamefont {N.}~\bibnamefont
  {Tsuji}}, \bibinfo {author} {\bibfnamefont {P.}~\bibnamefont {Barmettler}},
  \bibinfo {author} {\bibfnamefont {H.}~\bibnamefont {Aoki}},\ and\ \bibinfo
  {author} {\bibfnamefont {P.}~\bibnamefont {Werner}},\ }\href
  {https://doi.org/10.1103/PhysRevB.90.075117} {\bibfield  {journal} {\bibinfo
  {journal} {Phys. Rev. B}\ }\textbf {\bibinfo {volume} {90}},\ \bibinfo
  {pages} {075117} (\bibinfo {year} {2014})}\BibitemShut {NoStop}%
\bibitem [{\citenamefont {Eckstein}\ and\ \citenamefont
  {Werner}(2016)}]{Eckstein2016}%
  \BibitemOpen
  \bibfield  {author} {\bibinfo {author} {\bibfnamefont {M.}~\bibnamefont
  {Eckstein}}\ and\ \bibinfo {author} {\bibfnamefont {P.}~\bibnamefont
  {Werner}},\ }\href {https://doi.org/10.1038/srep21235} {\bibfield  {journal}
  {\bibinfo  {journal} {Scientific Reports}\ }\textbf {\bibinfo {volume} {6}},\
  \bibinfo {pages} {21235} (\bibinfo {year} {2016})}\BibitemShut {NoStop}%
\bibitem [{\citenamefont {Bittner}\ \emph {et~al.}(2020)\citenamefont
  {Bittner}, \citenamefont {Gole\ifmmode~\check{z}\else \v{z}\fi{}},
  \citenamefont {Eckstein},\ and\ \citenamefont {Werner}}]{Bittner2020}%
  \BibitemOpen
  \bibfield  {author} {\bibinfo {author} {\bibfnamefont {N.}~\bibnamefont
  {Bittner}}, \bibinfo {author} {\bibfnamefont {D.}~\bibnamefont
  {Gole\ifmmode~\check{z}\else \v{z}\fi{}}}, \bibinfo {author} {\bibfnamefont
  {M.}~\bibnamefont {Eckstein}},\ and\ \bibinfo {author} {\bibfnamefont
  {P.}~\bibnamefont {Werner}},\ }\href
  {https://doi.org/10.1103/PhysRevB.101.085127} {\bibfield  {journal} {\bibinfo
   {journal} {Phys. Rev. B}\ }\textbf {\bibinfo {volume} {101}},\ \bibinfo
  {pages} {085127} (\bibinfo {year} {2020})}\BibitemShut {NoStop}%
\bibitem [{\citenamefont {Jani\ifmmode~\check{s}\else \v{s}\fi{}}\ and\
  \citenamefont {Augustinsk\'y}(2008)}]{Janis2008}%
  \BibitemOpen
  \bibfield  {author} {\bibinfo {author} {\bibfnamefont {V.}~\bibnamefont
  {Jani\ifmmode~\check{s}\else \v{s}\fi{}}}\ and\ \bibinfo {author}
  {\bibfnamefont {P.}~\bibnamefont {Augustinsk\'y}},\ }\href
  {https://doi.org/10.1103/PhysRevB.77.085106} {\bibfield  {journal} {\bibinfo
  {journal} {Phys. Rev. B}\ }\textbf {\bibinfo {volume} {77}},\ \bibinfo
  {pages} {085106} (\bibinfo {year} {2008})}\BibitemShut {NoStop}%
\bibitem [{\citenamefont {Jani\ifmmode~\check{s}\else \v{s}\fi{}}\ \emph
  {et~al.}(2017{\natexlab{a}})\citenamefont {Jani\ifmmode~\check{s}\else
  \v{s}\fi{}}, \citenamefont {Kauch},\ and\ \citenamefont
  {Pokorn\'y}}]{Janis2017}%
  \BibitemOpen
  \bibfield  {author} {\bibinfo {author} {\bibfnamefont {V.}~\bibnamefont
  {Jani\ifmmode~\check{s}\else \v{s}\fi{}}}, \bibinfo {author} {\bibfnamefont
  {A.}~\bibnamefont {Kauch}},\ and\ \bibinfo {author} {\bibfnamefont
  {V.}~\bibnamefont {Pokorn\'y}},\ }\href
  {https://doi.org/10.1103/PhysRevB.95.045108} {\bibfield  {journal} {\bibinfo
  {journal} {Phys. Rev. B}\ }\textbf {\bibinfo {volume} {95}},\ \bibinfo
  {pages} {045108} (\bibinfo {year} {2017}{\natexlab{a}})}\BibitemShut
  {NoStop}%
\bibitem [{\citenamefont {Jani\ifmmode~\check{s}\else \v{s}\fi{}}\ \emph
  {et~al.}(2017{\natexlab{b}})\citenamefont {Jani\ifmmode~\check{s}\else
  \v{s}\fi{}}, \citenamefont {Pokorn\'y},\ and\ \citenamefont
  {Kauch}}]{Janis2017b}%
  \BibitemOpen
  \bibfield  {author} {\bibinfo {author} {\bibfnamefont {V.}~\bibnamefont
  {Jani\ifmmode~\check{s}\else \v{s}\fi{}}}, \bibinfo {author} {\bibfnamefont
  {V.}~\bibnamefont {Pokorn\'y}},\ and\ \bibinfo {author} {\bibfnamefont
  {A.}~\bibnamefont {Kauch}},\ }\href
  {https://doi.org/10.1103/PhysRevB.95.165113} {\bibfield  {journal} {\bibinfo
  {journal} {Phys. Rev. B}\ }\textbf {\bibinfo {volume} {95}},\ \bibinfo
  {pages} {165113} (\bibinfo {year} {2017}{\natexlab{b}})}\BibitemShut
  {NoStop}%
\bibitem [{\citenamefont {Zalom}\ \emph {et~al.}(2018)\citenamefont {Zalom},
  \citenamefont {Pokorn{\'y}},\ and\ \citenamefont {Jani{\v s}}}]{Zalom2018}%
  \BibitemOpen
  \bibfield  {author} {\bibinfo {author} {\bibfnamefont {P.}~\bibnamefont
  {Zalom}}, \bibinfo {author} {\bibfnamefont {V.}~\bibnamefont {Pokorn{\'y}}},\
  and\ \bibinfo {author} {\bibfnamefont {V.}~\bibnamefont {Jani{\v s}}},\
  }\href {https://doi.org/https://doi.org/10.1016/j.physb.2017.07.068}
  {\bibfield  {journal} {\bibinfo  {journal} {Physica B: Condensed Matter}\
  }\textbf {\bibinfo {volume} {536}},\ \bibinfo {pages} {704} (\bibinfo {year}
  {2018})}\BibitemShut {NoStop}%
\bibitem [{\citenamefont {Vilk}\ and\ \citenamefont
  {Tremblay}(1997)}]{Vilk1997}%
  \BibitemOpen
  \bibfield  {author} {\bibinfo {author} {\bibfnamefont {Y.}~\bibnamefont
  {Vilk}}\ and\ \bibinfo {author} {\bibfnamefont {A.-M.}\ \bibnamefont
  {Tremblay}},\ }\href@noop {} {\bibfield  {journal} {\bibinfo  {journal}
  {Journal de Physique I}\ }\textbf {\bibinfo {volume} {7}},\ \bibinfo {pages}
  {1309} (\bibinfo {year} {1997})}\BibitemShut {NoStop}%
\bibitem [{\citenamefont {Tremblay}(2012)}]{Tremblay2012}%
  \BibitemOpen
  \bibfield  {author} {\bibinfo {author} {\bibfnamefont {A.-M.~S.}\
  \bibnamefont {Tremblay}},\ }\bibinfo {title} {Two-particle-self-consistent
  approach for the hubbard model},\ in\ \href
  {https://doi.org/10.1007/978-3-642-21831-6_13} {\emph {\bibinfo {booktitle}
  {Strongly Correlated Systems: Theoretical Methods}}},\ \bibinfo {editor}
  {edited by\ \bibinfo {editor} {\bibfnamefont {A.}~\bibnamefont {Avella}}\
  and\ \bibinfo {editor} {\bibfnamefont {F.}~\bibnamefont {Mancini}}}\
  (\bibinfo  {publisher} {Springer Berlin Heidelberg},\ \bibinfo {address}
  {Berlin, Heidelberg},\ \bibinfo {year} {2012})\ pp.\ \bibinfo {pages}
  {409--453}\BibitemShut {NoStop}%
\bibitem [{\citenamefont {Kadanoff}\ and\ \citenamefont
  {Baym}(1962)}]{Kadanoff1962}%
  \BibitemOpen
  \bibfield  {author} {\bibinfo {author} {\bibfnamefont {L.~P.}\ \bibnamefont
  {Kadanoff}}\ and\ \bibinfo {author} {\bibfnamefont {G.~A.}\ \bibnamefont
  {Baym}},\ }\href@noop {} {\emph {\bibinfo {title} {Quantum statistical
  mechanics}}}\ (\bibinfo  {publisher} {Benjamin},\ \bibinfo {year}
  {1962})\BibitemShut {NoStop}%
\bibitem [{\citenamefont {Bruus}\ and\ \citenamefont
  {Flensberg}(2004)}]{Bruus2004}%
  \BibitemOpen
  \bibfield  {author} {\bibinfo {author} {\bibfnamefont {H.}~\bibnamefont
  {Bruus}}\ and\ \bibinfo {author} {\bibfnamefont {K.}~\bibnamefont
  {Flensberg}},\ }\href@noop {} {\emph {\bibinfo {title} {Many-body quantum
  theory in condensed matter physics: an introduction}}}\ (\bibinfo
  {publisher} {Oxford university press},\ \bibinfo {year} {2004})\BibitemShut
  {NoStop}%
\bibitem [{\citenamefont {Held}\ \emph {et~al.}(2008)\citenamefont {Held},
  \citenamefont {Katanin},\ and\ \citenamefont {Toschi}}]{Held2008}%
  \BibitemOpen
  \bibfield  {author} {\bibinfo {author} {\bibfnamefont {K.}~\bibnamefont
  {Held}}, \bibinfo {author} {\bibfnamefont {A.~A.}\ \bibnamefont {Katanin}},\
  and\ \bibinfo {author} {\bibfnamefont {A.}~\bibnamefont {Toschi}},\ }\href
  {https://doi.org/10.1143/PTPS.176.117} {\bibfield  {journal} {\bibinfo
  {journal} {Progress of Theoretical Physics Supplement}\ }\textbf {\bibinfo
  {volume} {176}},\ \bibinfo {pages} {117} (\bibinfo {year}
  {2008})}\BibitemShut {NoStop}%
\bibitem [{\citenamefont {Rubtsov}\ \emph {et~al.}(2008)\citenamefont
  {Rubtsov}, \citenamefont {Katsnelson},\ and\ \citenamefont
  {Lichtenstein}}]{Rubtsov2008}%
  \BibitemOpen
  \bibfield  {author} {\bibinfo {author} {\bibfnamefont {A.~N.}\ \bibnamefont
  {Rubtsov}}, \bibinfo {author} {\bibfnamefont {M.~I.}\ \bibnamefont
  {Katsnelson}},\ and\ \bibinfo {author} {\bibfnamefont {A.~I.}\ \bibnamefont
  {Lichtenstein}},\ }\href {https://doi.org/10.1103/PhysRevB.77.033101}
  {\bibfield  {journal} {\bibinfo  {journal} {Phys. Rev. B}\ }\textbf {\bibinfo
  {volume} {77}},\ \bibinfo {pages} {033101} (\bibinfo {year}
  {2008})}\BibitemShut {NoStop}%
\bibitem [{\citenamefont {Vilk}\ \emph
  {et~al.}(1994{\natexlab{a}})\citenamefont {Vilk}, \citenamefont {Chen},\ and\
  \citenamefont {Tremblay}}]{Vilk1994}%
  \BibitemOpen
  \bibfield  {author} {\bibinfo {author} {\bibfnamefont {Y.~M.}\ \bibnamefont
  {Vilk}}, \bibinfo {author} {\bibfnamefont {L.}~\bibnamefont {Chen}},\ and\
  \bibinfo {author} {\bibfnamefont {A.-M.~S.}\ \bibnamefont {Tremblay}},\
  }\href {https://doi.org/10.1103/PhysRevB.49.13267} {\bibfield  {journal}
  {\bibinfo  {journal} {Phys. Rev. B}\ }\textbf {\bibinfo {volume} {49}},\
  \bibinfo {pages} {13267} (\bibinfo {year} {1994}{\natexlab{a}})}\BibitemShut
  {NoStop}%
\bibitem [{\citenamefont {Davoudi}\ and\ \citenamefont
  {Tremblay}(2007)}]{Davoudi2007}%
  \BibitemOpen
  \bibfield  {author} {\bibinfo {author} {\bibfnamefont {B.}~\bibnamefont
  {Davoudi}}\ and\ \bibinfo {author} {\bibfnamefont {A.-M.~S.}\ \bibnamefont
  {Tremblay}},\ }\href {https://doi.org/10.1103/PhysRevB.76.085115} {\bibfield
  {journal} {\bibinfo  {journal} {Phys. Rev. B}\ }\textbf {\bibinfo {volume}
  {76}},\ \bibinfo {pages} {085115} (\bibinfo {year} {2007})}\BibitemShut
  {NoStop}%
\bibitem [{\citenamefont {Lessnich}\ \emph {et~al.}(2023)\citenamefont
  {Lessnich}, \citenamefont {Gauvin-Ndiaye}, \citenamefont {Valentí},\ and\
  \citenamefont {Tremblay}}]{Lessnich2023}%
  \BibitemOpen
  \bibfield  {author} {\bibinfo {author} {\bibfnamefont {D.}~\bibnamefont
  {Lessnich}}, \bibinfo {author} {\bibfnamefont {C.}~\bibnamefont
  {Gauvin-Ndiaye}}, \bibinfo {author} {\bibfnamefont {R.}~\bibnamefont
  {Valentí}},\ and\ \bibinfo {author} {\bibfnamefont {A.~M.~S.}\ \bibnamefont
  {Tremblay}},\ }\href@noop {} {\  (\bibinfo {year} {2023})},\ \Eprint
  {https://arxiv.org/abs/2307.15652} {arXiv:2307.15652 [cond-mat.str-el]}
  \BibitemShut {NoStop}%
\bibitem [{\citenamefont {Miyahara}\ \emph {et~al.}(2013)\citenamefont
  {Miyahara}, \citenamefont {Arita},\ and\ \citenamefont
  {Ikeda}}]{Miyahara2013}%
  \BibitemOpen
  \bibfield  {author} {\bibinfo {author} {\bibfnamefont {H.}~\bibnamefont
  {Miyahara}}, \bibinfo {author} {\bibfnamefont {R.}~\bibnamefont {Arita}},\
  and\ \bibinfo {author} {\bibfnamefont {H.}~\bibnamefont {Ikeda}},\ }\href
  {https://doi.org/10.1103/PhysRevB.87.045113} {\bibfield  {journal} {\bibinfo
  {journal} {Phys. Rev. B}\ }\textbf {\bibinfo {volume} {87}},\ \bibinfo
  {pages} {045113} (\bibinfo {year} {2013})}\BibitemShut {NoStop}%
\bibitem [{\citenamefont {Zantout}\ \emph {et~al.}(2021)\citenamefont
  {Zantout}, \citenamefont {Backes},\ and\ \citenamefont
  {Valent{\'\i}}}]{Zantout2021}%
  \BibitemOpen
  \bibfield  {author} {\bibinfo {author} {\bibfnamefont {K.}~\bibnamefont
  {Zantout}}, \bibinfo {author} {\bibfnamefont {S.}~\bibnamefont {Backes}},\
  and\ \bibinfo {author} {\bibfnamefont {R.}~\bibnamefont {Valent{\'\i}}},\
  }\href {https://doi.org/https://doi.org/10.1002/andp.202000399} {\bibfield
  {journal} {\bibinfo  {journal} {Annalen der Physik}\ }\textbf {\bibinfo
  {volume} {533}},\ \bibinfo {pages} {2000399} (\bibinfo {year}
  {2021})}\BibitemShut {NoStop}%
\bibitem [{\citenamefont {Gauvin-Ndiaye}\ \emph
  {et~al.}(2023{\natexlab{a}})\citenamefont {Gauvin-Ndiaye}, \citenamefont
  {Leblanc}, \citenamefont {Marin}, \citenamefont {Martin}, \citenamefont
  {Lessnich},\ and\ \citenamefont {Tremblay}}]{GauvinNdiaye2023}%
  \BibitemOpen
  \bibfield  {author} {\bibinfo {author} {\bibfnamefont {C.}~\bibnamefont
  {Gauvin-Ndiaye}}, \bibinfo {author} {\bibfnamefont {J.}~\bibnamefont
  {Leblanc}}, \bibinfo {author} {\bibfnamefont {S.}~\bibnamefont {Marin}},
  \bibinfo {author} {\bibfnamefont {N.}~\bibnamefont {Martin}}, \bibinfo
  {author} {\bibfnamefont {D.}~\bibnamefont {Lessnich}},\ and\ \bibinfo
  {author} {\bibfnamefont {A.~M.~S.}\ \bibnamefont {Tremblay}},\ }\href@noop {}
  {\  (\bibinfo {year} {2023}{\natexlab{a}})},\ \Eprint
  {https://arxiv.org/abs/2308.14091} {arXiv:2308.14091 [cond-mat.str-el]}
  \BibitemShut {NoStop}%
\bibitem [{\citenamefont {Gauvin-Ndiaye}\ \emph
  {et~al.}(2023{\natexlab{b}})\citenamefont {Gauvin-Ndiaye}, \citenamefont
  {Lahaie}, \citenamefont {Vilk},\ and\ \citenamefont
  {Tremblay}}]{GauvinNdiaye2023a}%
  \BibitemOpen
  \bibfield  {author} {\bibinfo {author} {\bibfnamefont {C.}~\bibnamefont
  {Gauvin-Ndiaye}}, \bibinfo {author} {\bibfnamefont {C.}~\bibnamefont
  {Lahaie}}, \bibinfo {author} {\bibfnamefont {Y.~M.}\ \bibnamefont {Vilk}},\
  and\ \bibinfo {author} {\bibfnamefont {A.-M.~S.}\ \bibnamefont {Tremblay}},\
  }\href {https://doi.org/10.1103/PhysRevB.108.075144} {\bibfield  {journal}
  {\bibinfo  {journal} {Phys. Rev. B}\ }\textbf {\bibinfo {volume} {108}},\
  \bibinfo {pages} {075144} (\bibinfo {year} {2023}{\natexlab{b}})}\BibitemShut
  {NoStop}%
\bibitem [{\citenamefont {Martin}\ \emph {et~al.}(2023)\citenamefont {Martin},
  \citenamefont {Gauvin-Ndiaye},\ and\ \citenamefont {Tremblay}}]{Martin2023}%
  \BibitemOpen
  \bibfield  {author} {\bibinfo {author} {\bibfnamefont {N.}~\bibnamefont
  {Martin}}, \bibinfo {author} {\bibfnamefont {C.}~\bibnamefont
  {Gauvin-Ndiaye}},\ and\ \bibinfo {author} {\bibfnamefont {A.-M.~S.}\
  \bibnamefont {Tremblay}},\ }\href
  {https://doi.org/10.1103/PhysRevB.107.075158} {\bibfield  {journal} {\bibinfo
   {journal} {Phys. Rev. B}\ }\textbf {\bibinfo {volume} {107}},\ \bibinfo
  {pages} {075158} (\bibinfo {year} {2023})}\BibitemShut {NoStop}%
\bibitem [{\citenamefont {Zantout}\ \emph {et~al.}(2023)\citenamefont
  {Zantout}, \citenamefont {Backes}, \citenamefont {Razpopov}, \citenamefont
  {Lessnich},\ and\ \citenamefont {Valent\'{\i}}}]{Zantout2023}%
  \BibitemOpen
  \bibfield  {author} {\bibinfo {author} {\bibfnamefont {K.}~\bibnamefont
  {Zantout}}, \bibinfo {author} {\bibfnamefont {S.}~\bibnamefont {Backes}},
  \bibinfo {author} {\bibfnamefont {A.}~\bibnamefont {Razpopov}}, \bibinfo
  {author} {\bibfnamefont {D.}~\bibnamefont {Lessnich}},\ and\ \bibinfo
  {author} {\bibfnamefont {R.}~\bibnamefont {Valent\'{\i}}},\ }\href
  {https://doi.org/10.1103/PhysRevB.107.235101} {\bibfield  {journal} {\bibinfo
   {journal} {Phys. Rev. B}\ }\textbf {\bibinfo {volume} {107}},\ \bibinfo
  {pages} {235101} (\bibinfo {year} {2023})}\BibitemShut {NoStop}%
\bibitem [{\citenamefont {Simard}\ and\ \citenamefont
  {Werner}(2023)}]{Simard2023}%
  \BibitemOpen
  \bibfield  {author} {\bibinfo {author} {\bibfnamefont {O.}~\bibnamefont
  {Simard}}\ and\ \bibinfo {author} {\bibfnamefont {P.}~\bibnamefont
  {Werner}},\ }\href {https://doi.org/10.1103/PhysRevB.107.245137} {\bibfield
  {journal} {\bibinfo  {journal} {Phys. Rev. B}\ }\textbf {\bibinfo {volume}
  {107}},\ \bibinfo {pages} {245137} (\bibinfo {year} {2023})}\BibitemShut
  {NoStop}%
\bibitem [{\citenamefont {Yan}\ and\ \citenamefont {Jani\ifmmode~\check{s}\else
  \v{s}\fi{}}(2022)}]{Yan2022}%
  \BibitemOpen
  \bibfield  {author} {\bibinfo {author} {\bibfnamefont {J.}~\bibnamefont
  {Yan}}\ and\ \bibinfo {author} {\bibfnamefont {V.}~\bibnamefont
  {Jani\ifmmode~\check{s}\else \v{s}\fi{}}},\ }\href
  {https://doi.org/10.1103/PhysRevB.105.085122} {\bibfield  {journal} {\bibinfo
   {journal} {Phys. Rev. B}\ }\textbf {\bibinfo {volume} {105}},\ \bibinfo
  {pages} {085122} (\bibinfo {year} {2022})}\BibitemShut {NoStop}%
\bibitem [{\citenamefont {Simard}\ and\ \citenamefont
  {Werner}(2022)}]{Simard2022}%
  \BibitemOpen
  \bibfield  {author} {\bibinfo {author} {\bibfnamefont {O.}~\bibnamefont
  {Simard}}\ and\ \bibinfo {author} {\bibfnamefont {P.}~\bibnamefont
  {Werner}},\ }\href {https://doi.org/10.1103/PhysRevB.106.L241110} {\bibfield
  {journal} {\bibinfo  {journal} {Phys. Rev. B}\ }\textbf {\bibinfo {volume}
  {106}},\ \bibinfo {pages} {L241110} (\bibinfo {year} {2022})}\BibitemShut
  {NoStop}%
\bibitem [{\citenamefont {Karakuzu}\ \emph {et~al.}(2021)\citenamefont
  {Karakuzu}, \citenamefont {Johnston},\ and\ \citenamefont
  {Maier}}]{Karakuzu2021}%
  \BibitemOpen
  \bibfield  {author} {\bibinfo {author} {\bibfnamefont {S.}~\bibnamefont
  {Karakuzu}}, \bibinfo {author} {\bibfnamefont {S.}~\bibnamefont {Johnston}},\
  and\ \bibinfo {author} {\bibfnamefont {T.~A.}\ \bibnamefont {Maier}},\ }\href
  {https://doi.org/10.1103/PhysRevB.104.245109} {\bibfield  {journal} {\bibinfo
   {journal} {Phys. Rev. B}\ }\textbf {\bibinfo {volume} {104}},\ \bibinfo
  {pages} {245109} (\bibinfo {year} {2021})}\BibitemShut {NoStop}%
\bibitem [{\citenamefont {Yue}\ \emph {et~al.}(2022)\citenamefont {Yue},
  \citenamefont {Aoki},\ and\ \citenamefont {Werner}}]{Yue2022}%
  \BibitemOpen
  \bibfield  {author} {\bibinfo {author} {\bibfnamefont {C.}~\bibnamefont
  {Yue}}, \bibinfo {author} {\bibfnamefont {H.}~\bibnamefont {Aoki}},\ and\
  \bibinfo {author} {\bibfnamefont {P.}~\bibnamefont {Werner}},\ }\href
  {https://doi.org/10.1103/PhysRevB.106.L180506} {\bibfield  {journal}
  {\bibinfo  {journal} {Phys. Rev. B}\ }\textbf {\bibinfo {volume} {106}},\
  \bibinfo {pages} {L180506} (\bibinfo {year} {2022})}\BibitemShut {NoStop}%
\bibitem [{\citenamefont {Zeng}\ \emph {et~al.}(2023)\citenamefont {Zeng},
  \citenamefont {Crépel},\ and\ \citenamefont {Millis}}]{Zeng2023}%
  \BibitemOpen
  \bibfield  {author} {\bibinfo {author} {\bibfnamefont {Y.}~\bibnamefont
  {Zeng}}, \bibinfo {author} {\bibfnamefont {V.}~\bibnamefont {Crépel}},\ and\
  \bibinfo {author} {\bibfnamefont {A.~J.}\ \bibnamefont {Millis}},\
  }\href@noop {} {\  (\bibinfo {year} {2023})},\ \Eprint
  {https://arxiv.org/abs/2311.04074} {arXiv:2311.04074 [cond-mat.mes-hall]}
  \BibitemShut {NoStop}%
\bibitem [{\citenamefont {Stefanucci}\ and\ \citenamefont
  {Van~Leeuwen}(2013)}]{stefanucci2013nonequilibrium}%
  \BibitemOpen
  \bibfield  {author} {\bibinfo {author} {\bibfnamefont {G.}~\bibnamefont
  {Stefanucci}}\ and\ \bibinfo {author} {\bibfnamefont {R.}~\bibnamefont
  {Van~Leeuwen}},\ }\href@noop {} {\emph {\bibinfo {title} {Nonequilibrium
  many-body theory of quantum systems: a modern introduction}}}\ (\bibinfo
  {publisher} {Cambridge University Press},\ \bibinfo {year}
  {2013})\BibitemShut {NoStop}%
\bibitem [{\citenamefont {Haug}\ \emph {et~al.}(2008)\citenamefont {Haug},
  \citenamefont {Jauho},\ and\ \citenamefont {Cardona}}]{Haug2008}%
  \BibitemOpen
  \bibfield  {author} {\bibinfo {author} {\bibfnamefont {H.}~\bibnamefont
  {Haug}}, \bibinfo {author} {\bibfnamefont {A.-P.}\ \bibnamefont {Jauho}},\
  and\ \bibinfo {author} {\bibfnamefont {M.}~\bibnamefont {Cardona}},\
  }\href@noop {} {\emph {\bibinfo {title} {Quantum kinetics in transport and
  optics of semiconductors}}},\ Vol.~\bibinfo {volume} {2}\ (\bibinfo
  {publisher} {Springer},\ \bibinfo {year} {2008})\BibitemShut {NoStop}%
\bibitem [{Note1()}]{Note1}%
  \BibitemOpen
  \bibinfo {note} {$2 \rightarrow 1^-$ leads to same results, since $\chi
  ^{\protect \text {sp/ch}}(1,1^+) = \chi ^{\protect \text
  {sp/ch}}(1,1^-)$.}\BibitemShut {Stop}%
\bibitem [{\citenamefont {Vilk}\ \emph
  {et~al.}(1994{\natexlab{b}})\citenamefont {Vilk}, \citenamefont {Chen},\ and\
  \citenamefont {Tremblay}}]{Vilk1994a}%
  \BibitemOpen
  \bibfield  {author} {\bibinfo {author} {\bibfnamefont {Y.}~\bibnamefont
  {Vilk}}, \bibinfo {author} {\bibfnamefont {L.}~\bibnamefont {Chen}},\ and\
  \bibinfo {author} {\bibfnamefont {A.-M.}\ \bibnamefont {Tremblay}},\ }\href
  {https://doi.org/https://doi.org/10.1016/0921-4534(94)92339-6} {\bibfield
  {journal} {\bibinfo  {journal} {Physica C: Superconductivity}\ }\textbf
  {\bibinfo {volume} {235-240}},\ \bibinfo {pages} {2235} (\bibinfo {year}
  {1994}{\natexlab{b}})}\BibitemShut {NoStop}%
\bibitem [{\citenamefont {Allen}\ \emph {et~al.}(2004)\citenamefont {Allen},
  \citenamefont {Tremblay},\ and\ \citenamefont {Vilk}}]{Allen2004}%
  \BibitemOpen
  \bibfield  {author} {\bibinfo {author} {\bibfnamefont {S.}~\bibnamefont
  {Allen}}, \bibinfo {author} {\bibfnamefont {A.~M.~S.}\ \bibnamefont
  {Tremblay}},\ and\ \bibinfo {author} {\bibfnamefont {Y.~M.}\ \bibnamefont
  {Vilk}},\ }\bibinfo {title} {Conserving approximations vs. two-particle
  self-consistent approach},\ in\ \href
  {https://doi.org/10.1007/0-387-21717-7_8} {\emph {\bibinfo {booktitle}
  {Theoretical Methods for Strongly Correlated Electrons}}},\ \bibinfo {editor}
  {edited by\ \bibinfo {editor} {\bibfnamefont {D.}~\bibnamefont
  {S{\'e}n{\'e}chal}}, \bibinfo {editor} {\bibfnamefont {A.-M.}\ \bibnamefont
  {Tremblay}},\ and\ \bibinfo {editor} {\bibfnamefont {C.}~\bibnamefont
  {Bourbonnais}}}\ (\bibinfo  {publisher} {Springer New York},\ \bibinfo
  {address} {New York, NY},\ \bibinfo {year} {2004})\ pp.\ \bibinfo {pages}
  {341--355}\BibitemShut {NoStop}%
\bibitem [{\citenamefont {Yan}\ and\ \citenamefont {Werner}(2023)}]{Yan2023}%
  \BibitemOpen
  \bibfield  {author} {\bibinfo {author} {\bibfnamefont {J.}~\bibnamefont
  {Yan}}\ and\ \bibinfo {author} {\bibfnamefont {P.}~\bibnamefont {Werner}},\
  }\href {https://doi.org/10.1103/PhysRevB.108.125143} {\bibfield  {journal}
  {\bibinfo  {journal} {Phys. Rev. B}\ }\textbf {\bibinfo {volume} {108}},\
  \bibinfo {pages} {125143} (\bibinfo {year} {2023})}\BibitemShut {NoStop}%
\bibitem [{\citenamefont {Yan}\ and\ \citenamefont {Ke}(2016)}]{Yan2016}%
  \BibitemOpen
  \bibfield  {author} {\bibinfo {author} {\bibfnamefont {J.}~\bibnamefont
  {Yan}}\ and\ \bibinfo {author} {\bibfnamefont {Y.}~\bibnamefont {Ke}},\
  }\href {https://doi.org/10.1103/PhysRevB.94.045424} {\bibfield  {journal}
  {\bibinfo  {journal} {Phys. Rev. B}\ }\textbf {\bibinfo {volume} {94}},\
  \bibinfo {pages} {045424} (\bibinfo {year} {2016})}\BibitemShut {NoStop}%
\bibitem [{\citenamefont {Datta}(2005)}]{Datta2005}%
  \BibitemOpen
  \bibfield  {author} {\bibinfo {author} {\bibfnamefont {S.}~\bibnamefont
  {Datta}},\ }\href@noop {} {\emph {\bibinfo {title} {Quantum transport: atom
  to transistor}}}\ (\bibinfo  {publisher} {Cambridge University Press},\
  \bibinfo {year} {2005})\BibitemShut {NoStop}%
\bibitem [{\citenamefont {Klett}\ \emph {et~al.}(2020)\citenamefont {Klett},
  \citenamefont {Wentzell}, \citenamefont {Sch\"afer}, \citenamefont
  {Simkovic}, \citenamefont {Parcollet}, \citenamefont {Andergassen},\ and\
  \citenamefont {Hansmann}}]{PhysRevResearch.2.033476}%
  \BibitemOpen
  \bibfield  {author} {\bibinfo {author} {\bibfnamefont {M.}~\bibnamefont
  {Klett}}, \bibinfo {author} {\bibfnamefont {N.}~\bibnamefont {Wentzell}},
  \bibinfo {author} {\bibfnamefont {T.}~\bibnamefont {Sch\"afer}}, \bibinfo
  {author} {\bibfnamefont {F.}~\bibnamefont {Simkovic}}, \bibinfo {author}
  {\bibfnamefont {O.}~\bibnamefont {Parcollet}}, \bibinfo {author}
  {\bibfnamefont {S.}~\bibnamefont {Andergassen}},\ and\ \bibinfo {author}
  {\bibfnamefont {P.}~\bibnamefont {Hansmann}},\ }\href
  {https://doi.org/10.1103/PhysRevResearch.2.033476} {\bibfield  {journal}
  {\bibinfo  {journal} {Phys. Rev. Res.}\ }\textbf {\bibinfo {volume} {2}},\
  \bibinfo {pages} {033476} (\bibinfo {year} {2020})}\BibitemShut {NoStop}%
\bibitem [{\citenamefont {Brener}\ \emph {et~al.}(2008)\citenamefont {Brener},
  \citenamefont {Hafermann}, \citenamefont {Rubtsov}, \citenamefont
  {Katsnelson},\ and\ \citenamefont {Lichtenstein}}]{Brener2008}%
  \BibitemOpen
  \bibfield  {author} {\bibinfo {author} {\bibfnamefont {S.}~\bibnamefont
  {Brener}}, \bibinfo {author} {\bibfnamefont {H.}~\bibnamefont {Hafermann}},
  \bibinfo {author} {\bibfnamefont {A.~N.}\ \bibnamefont {Rubtsov}}, \bibinfo
  {author} {\bibfnamefont {M.~I.}\ \bibnamefont {Katsnelson}},\ and\ \bibinfo
  {author} {\bibfnamefont {A.~I.}\ \bibnamefont {Lichtenstein}},\ }\href
  {https://doi.org/10.1103/PhysRevB.77.195105} {\bibfield  {journal} {\bibinfo
  {journal} {Phys. Rev. B}\ }\textbf {\bibinfo {volume} {77}},\ \bibinfo
  {pages} {195105} (\bibinfo {year} {2008})}\BibitemShut {NoStop}%
\bibitem [{\citenamefont {Golor}\ \emph {et~al.}(2014)\citenamefont {Golor},
  \citenamefont {Reckling}, \citenamefont {Classen}, \citenamefont {Scherer},\
  and\ \citenamefont {Wessel}}]{Golor2014}%
  \BibitemOpen
  \bibfield  {author} {\bibinfo {author} {\bibfnamefont {M.}~\bibnamefont
  {Golor}}, \bibinfo {author} {\bibfnamefont {T.}~\bibnamefont {Reckling}},
  \bibinfo {author} {\bibfnamefont {L.}~\bibnamefont {Classen}}, \bibinfo
  {author} {\bibfnamefont {M.~M.}\ \bibnamefont {Scherer}},\ and\ \bibinfo
  {author} {\bibfnamefont {S.}~\bibnamefont {Wessel}},\ }\href
  {https://doi.org/10.1103/PhysRevB.90.195131} {\bibfield  {journal} {\bibinfo
  {journal} {Phys. Rev. B}\ }\textbf {\bibinfo {volume} {90}},\ \bibinfo
  {pages} {195131} (\bibinfo {year} {2014})}\BibitemShut {NoStop}%
\bibitem [{\citenamefont {Gukelberger}\ \emph {et~al.}(2015)\citenamefont
  {Gukelberger}, \citenamefont {Huang},\ and\ \citenamefont
  {Werner}}]{Gukelberger2015}%
  \BibitemOpen
  \bibfield  {author} {\bibinfo {author} {\bibfnamefont {J.}~\bibnamefont
  {Gukelberger}}, \bibinfo {author} {\bibfnamefont {L.}~\bibnamefont {Huang}},\
  and\ \bibinfo {author} {\bibfnamefont {P.}~\bibnamefont {Werner}},\ }\href
  {https://doi.org/10.1103/PhysRevB.91.235114} {\bibfield  {journal} {\bibinfo
  {journal} {Phys. Rev. B}\ }\textbf {\bibinfo {volume} {91}},\ \bibinfo
  {pages} {235114} (\bibinfo {year} {2015})}\BibitemShut {NoStop}%
\bibitem [{\citenamefont {Eckstein}\ \emph {et~al.}(2010)\citenamefont
  {Eckstein}, \citenamefont {Kollar},\ and\ \citenamefont
  {Werner}}]{Eckstein2010}%
  \BibitemOpen
  \bibfield  {author} {\bibinfo {author} {\bibfnamefont {M.}~\bibnamefont
  {Eckstein}}, \bibinfo {author} {\bibfnamefont {M.}~\bibnamefont {Kollar}},\
  and\ \bibinfo {author} {\bibfnamefont {P.}~\bibnamefont {Werner}},\ }\href
  {https://doi.org/10.1103/PhysRevB.81.115131} {\bibfield  {journal} {\bibinfo
  {journal} {Phys. Rev. B}\ }\textbf {\bibinfo {volume} {81}},\ \bibinfo
  {pages} {115131} (\bibinfo {year} {2010})}\BibitemShut {NoStop}%
\bibitem [{\citenamefont {Gall}\ \emph {et~al.}(2021)\citenamefont {Gall},
  \citenamefont {Wurz}, \citenamefont {Samland}, \citenamefont {Chan},\ and\
  \citenamefont {K{\"o}hl}}]{Gall2021}%
  \BibitemOpen
  \bibfield  {author} {\bibinfo {author} {\bibfnamefont {M.}~\bibnamefont
  {Gall}}, \bibinfo {author} {\bibfnamefont {N.}~\bibnamefont {Wurz}}, \bibinfo
  {author} {\bibfnamefont {J.}~\bibnamefont {Samland}}, \bibinfo {author}
  {\bibfnamefont {C.~F.}\ \bibnamefont {Chan}},\ and\ \bibinfo {author}
  {\bibfnamefont {M.}~\bibnamefont {K{\"o}hl}},\ }\href
  {https://doi.org/10.1038/s41586-020-03058-x} {\bibfield  {journal} {\bibinfo
  {journal} {Nature}\ }\textbf {\bibinfo {volume} {589}},\ \bibinfo {pages}
  {40} (\bibinfo {year} {2021})}\BibitemShut {NoStop}%
\bibitem [{\citenamefont {Trotzky}\ \emph {et~al.}(2008)\citenamefont
  {Trotzky}, \citenamefont {Cheinet}, \citenamefont {F{\"o}lling},
  \citenamefont {Feld}, \citenamefont {Schnorrberger}, \citenamefont {Rey},
  \citenamefont {Polkovnikov}, \citenamefont {Demler}, \citenamefont {Lukin},\
  and\ \citenamefont {Bloch}}]{Trotzky2008}%
  \BibitemOpen
  \bibfield  {author} {\bibinfo {author} {\bibfnamefont {S.}~\bibnamefont
  {Trotzky}}, \bibinfo {author} {\bibfnamefont {P.}~\bibnamefont {Cheinet}},
  \bibinfo {author} {\bibfnamefont {S.}~\bibnamefont {F{\"o}lling}}, \bibinfo
  {author} {\bibfnamefont {M.}~\bibnamefont {Feld}}, \bibinfo {author}
  {\bibfnamefont {U.}~\bibnamefont {Schnorrberger}}, \bibinfo {author}
  {\bibfnamefont {A.~M.}\ \bibnamefont {Rey}}, \bibinfo {author} {\bibfnamefont
  {A.}~\bibnamefont {Polkovnikov}}, \bibinfo {author} {\bibfnamefont {E.~A.}\
  \bibnamefont {Demler}}, \bibinfo {author} {\bibfnamefont {M.~D.}\
  \bibnamefont {Lukin}},\ and\ \bibinfo {author} {\bibfnamefont
  {I.}~\bibnamefont {Bloch}},\ }\href {https://doi.org/10.1126/science.1150841}
  {\bibfield  {journal} {\bibinfo  {journal} {Science}\ }\textbf {\bibinfo
  {volume} {319}},\ \bibinfo {pages} {295} (\bibinfo {year}
  {2008})}\BibitemShut {NoStop}%
\bibitem [{\citenamefont {Dasari}\ and\ \citenamefont
  {Eckstein}(2019)}]{Dasari2019}%
  \BibitemOpen
  \bibfield  {author} {\bibinfo {author} {\bibfnamefont {N.}~\bibnamefont
  {Dasari}}\ and\ \bibinfo {author} {\bibfnamefont {M.}~\bibnamefont
  {Eckstein}},\ }\href {https://doi.org/10.1103/PhysRevB.100.121114} {\bibfield
   {journal} {\bibinfo  {journal} {Phys. Rev. B}\ }\textbf {\bibinfo {volume}
  {100}},\ \bibinfo {pages} {121114} (\bibinfo {year} {2019})}\BibitemShut
  {NoStop}%
\bibitem [{\citenamefont {Loh\"ofer}\ \emph {et~al.}(2015)\citenamefont
  {Loh\"ofer}, \citenamefont {Coletta}, \citenamefont {Joshi}, \citenamefont
  {Assaad}, \citenamefont {Vojta}, \citenamefont {Wessel},\ and\ \citenamefont
  {Mila}}]{Lohoefer2015}%
  \BibitemOpen
  \bibfield  {author} {\bibinfo {author} {\bibfnamefont {M.}~\bibnamefont
  {Loh\"ofer}}, \bibinfo {author} {\bibfnamefont {T.}~\bibnamefont {Coletta}},
  \bibinfo {author} {\bibfnamefont {D.~G.}\ \bibnamefont {Joshi}}, \bibinfo
  {author} {\bibfnamefont {F.~F.}\ \bibnamefont {Assaad}}, \bibinfo {author}
  {\bibfnamefont {M.}~\bibnamefont {Vojta}}, \bibinfo {author} {\bibfnamefont
  {S.}~\bibnamefont {Wessel}},\ and\ \bibinfo {author} {\bibfnamefont
  {F.}~\bibnamefont {Mila}},\ }\href
  {https://doi.org/10.1103/PhysRevB.92.245137} {\bibfield  {journal} {\bibinfo
  {journal} {Phys. Rev. B}\ }\textbf {\bibinfo {volume} {92}},\ \bibinfo
  {pages} {245137} (\bibinfo {year} {2015})}\BibitemShut {NoStop}%
\end{thebibliography}%

\end{document}